\documentstyle[]{mn}

\newlength{\colwidthf}
\newlength{\hcolwidthf}
\newlength{\colwidth}
\newlength{\hcolwidth}
\input ./psfig.sty
\newif\ifpsfiles\psfilestrue
\psfilestrue
\def\getfig#1#2{\ifpsfiles\psfig{figure=#1,width=\hsize}\else\vskip#2\fi}

\newcommand{\bea}{\begin{eqnarray*}}
\newcommand{\eea}{\end{eqnarray*}}
\newcommand{\be}{\begin{eqnarray}}
\newcommand{\ee}{\end{eqnarray}}

\def\phibar{\varphi_{\rm bar}}

\def\kpc{\,{\rm kpc}}

\def\etal{{\rm et al. }}

\def\los{{LOS}}

\def\spose#1{\hbox to 0pt{#1\hss}}
\def\lta{\mathrel{\spose{\lower 3pt\hbox{$\mathchar"218$}}
     \raise 2.0pt\hbox{$\mathchar"13C$}}}
\def\gta{\mathrel{\spose{\lower 3pt\hbox{$\mathchar"218$}}
     \raise 2.0pt\hbox{$\mathchar"13E$}}}
\def\pc{{\rm\,pc}}
\def\kpc{{\rm\,kpc}}
\def\msun{{\rm\,M_\odot}}

\title{Spiral arms, bar shape and bulge microlensing in the Milky Way}

\author[N. Bissantz and O. Gerhard]{Nicolai Bissantz$^1$ and Ortwin
Gerhard$^1$\\
$^1$Astronomisches Institut der Universit\"at Basel, Venusstr.7,
CH-4102 Binningen}

\begin{document}
\maketitle

\begin{abstract}
  
A new model for the luminosity distribution in the inner Milky Way is
found, using a non-parametric penalized maximum-likelihood algorithm
to deproject a dereddened {\sl COBE/DIRBE} L-band map of the inner
Galaxy. The model is also constrained by the apparent magnitude
(line-of-sight) distributions of clump giant stars in certain bulge
fields.  An important new feature is the inclusion of a spiral arm
model in the disk.

Spiral arms make the model appear broader on the sky, thus our bar is
more elongated than in previous eight-fold symmetric models.  They
also lead to a smoother disk model interior to the Sun.  The bar
length is $\approx\! 3.5\kpc$ and its axis ratios are 1:(0.3-0.4):0.3,
independent of whether the spiral arm model is 4-armed or 2-armed.
The larger elongation in the plane makes it possible to reproduce the
observed clump giant distributions as well. With only the surface
brightness data a small model degeneracy is found even for fixed
orientation of the bar, amounting to about $\pm 0.1$ uncertainty in
the in-plane axial ratio. Including the clump giant data removes most
of this degeneracy and also places additional constraints on the bar's
orientation angle. We estimate $15\deg \lta \phibar \lta 30\deg$, with
the best models obtained for $20\deg \lta \phibar \lta 25\deg$.

We use our reference model to predict a microlensing optical depth map
towards the bulge, normalising its mass by the observed terminal velocity
curve. For clump giant sources at $(l,b)=(3.9\deg,-3.8\deg)$ we find
$\tau_{-6}\equiv\tau/10^{-6}=1.27$, within $1.8\sigma$ of the new
MACHO measurement given by Popowski et al. The value for all sources
at $(l,b)=(2.68\deg,-3.35\deg)$ is $\tau_{-6}=1.1$, still $>3\sigma$
away from the published MACHO DIA value. The dispersion of these
$\tau_{-6}$ values within our models is $\simeq 10\%$. Because the
distribution of sources is well-fit by the NIR model, increasing the
predicted optical depths by $>20\%$ will be difficult.
Thus the high value of the measured clump giant optical depth argues
for a near-maximal disk in the Milky Way. 

\end{abstract}


\section{Introduction}
Observations of the Milky Way [MW] show significant systematic
differences between the NIR surface brightness of the MW at $l>0\deg$
and $l<0\deg$ (e.g., Blitz \& Spergel \cite{blitz91}, Weiland \etal
\cite{weiland94}, Bissantz \etal \cite{bissantz97}). It is widely accepted that these
variations reflect the fact that the MW is a barred spiral galaxy.
Evidence for a barred component of the luminosity density in the inner
MW also comes from starcount observations (e.g., Stanek \etal
\cite{stanek97}, Nikolaev \& Weinberg  \cite{nw97}, Sevenster 
\cite{sev99}, Hammersley \etal \cite{hamm99}, L\'opez-Corredoira \etal 
\cite{lop2000}), from gas-dynamics (e.g., Englmaier \& Gerhard 
\cite{ppe99}, Fux \cite{fux99}, Weiner \& Sellwood \cite{ws99}), and
microlensing observations (e.g., Paczynski \etal \cite{pac94}, Zhao,
Rich \& Spergel \cite{zrs96}). Further references can be found in
Gerhard (\cite{og01}).

The starcount data show significant asymmetries between lines-of-sight
that are symmetrical with respect to the $l=0$ axis; this is the
signature of a bar with its near end at positive Galactic longitudes.
Most importantly, starcount data contain information about the
distances to the surveyed stars. This is complementary to the all-sky
coverage of surface brightness maps, and is valuable for constraining
the line-of-sight structure of the bulge even if available only for a
restricted number of fields.  In this paper we will take one step
towards combining the information from both kinds of data, and use the
clump giant observations of Stanek \etal (\cite{stanek94},
\cite{stanek97}) together with the {\sl COBE/DIRBE} NIR data to
determine a model for the luminosity distribution in the inner Galaxy.
With this model we can be more confident about the line-of-sight
distribution (LOS) of microlensing sources, and are thus in a much better
position to predict microlensing optical depths for comparison with
the recent determinations from the MACHO group (Alcock \etal
\cite{alcock2000a}, Popowski \etal \cite{popo2000}).

Most previous models of the inner MW have been parametric and are thus
restricted towards certain classes of densities for the bulge and/or
disk.  Binney \& Gerhard \cite{bg96} developed a nonparametric
approach to the deprojection of the {\sl COBE/DIRBE} data based on the
Richardson-Lucy algorithm, in which by construction the luminosity
models are eight-fold symmetric with respect to the three main planes
of the bar/bulge.  Models constructed with this approach (Binney \etal
\cite{bgs97}; Bissantz \etal \cite{bissantz97}) give a good fit to the
{\sl COBE/DIRBE} L-band data, but predict less asymmetric {\los}
distributions towards the fields observed by Stanek \etal
\cite{stanek94} than observed, by more than $0.1^m$.  Eight-fold
symmetry also excludes modeling the spiral arms of the MW (see, e.g.,
Englmaier \& Gerhard \cite{ppe99}, Drimmel \& Spergel \cite{drimmsp01}). In
the present paper we describe a non-parametric penalized likelihood
approach to infer the luminosity density of the inner MW from the
{\sl COBE/DIRBE} data which allows us to include a spiral arm model.

This paper is organized as follows. Section 2 describes our new
deprojection algorithm. In Section 3 we test the method with known
parametric distributions and analyse the uniqueness of the deprojected
bar shape.  In Section 4 we present models for the luminosity
distribution of the MW which are consistent with both the {\sl
  COBE/DIRBE} L-band data and the observed asymmetry in the
distribution of clump giant stars, and give constraints on the
orientation angle of the Galactic bar. In Section 5 we predict the
microlensing optical depths for these models and compare to recently
published results of the MACHO experiment.  We close with a summary
and conclusions in Section 6.

\section{Maximum likelihood deprojection method}
\label{secmldmethod}

In this Section we describe the technique we have used to construct
models for the Milky Way's luminosity distribution. It is a
non-parametric technique that maximizes a likelihood function, which
includes penalty terms encouraging smoothness, eight-fold (triaxial)
symmetry and a spiral arm component in the model. The minimisation
procedure is iterative, starting from an initial parametric model. The
following subsections describe the initial parametric models
(\S\ref{secmldmethod}.1), the algorithm (\S\ref{secmldmethod}.2),
the choice of optimal penalty parameters (\S\ref{secmldmethod}.3) and
the performance of the algorithm (\S\ref{secmldmethod}.4). The
results of using the algorithm to recover known solutions from
artificial data are described in \S\ref{secseq}.

\subsection{Parametric models}
\label{secparametric}

We define parametric models for the luminosity distribution of the MW
on a cartesian grid.  The coordinate system has the Galactic centre at
its origin.  The axes are parallel to the main axes of the bar.  In
this coordinate system the position of the Sun is ($x\! =\!R_0\!
\cdot\!\cos (\phibar), y\! =\!R_0 \!\cdot\!\sin(\phibar), Z_0$), where
$R_0$ is the distance of the Sun from the Galactic centre projected
onto the main plane of the Milky Way, \(Z_0\) the position of the Sun
above the $xy$-plane, and $\phibar$ the ``bar angle'', i.e., the
angle in the $xy$-plane between the major axis of the bar and the
projected line-of-sight from the Sun to the Galactic centre, such
that for positive $\phibar$ the near end of the bar is at positive
longitudes. Throughout this paper we will use $R_0=8\kpc$ and $Z_0=14\pc$.

Our parametric models contain a double-exponential disk and a
truncated power-law bulge (cf. Binney, Gerhard \& Spergel \cite{bgs97}):
\be
\label{eqpara}
\hat{\rho}(\vec{x}) = \rho_{d}(\vec{x}) + \rho_{b}(\vec{x}),
\ee
where
\be
\rho_{d}\equiv \rho_d^0 \cdot R_d \cdot e^{-R/R_d} \cdot
\left(\frac{e^{-|z|/z_0}}{z_0}+\alpha \frac{e^{-|z|/z_1}}{z_1}\right),
\nonumber
\ee
\be
\rho_{b}\equiv \frac{\rho_b^0}{\eta\zeta a_m^3} \cdot \frac{e^{-a^2/a_m^2}}
{\left( 1+a/a_0 \right)^{1.8}},
\nonumber
\ee
\be
a\equiv \sqrt[]{x^2+\frac{y^2}{\eta^2} + \frac{z^2}{\zeta^2}}, \quad
R\equiv  \sqrt[]{x^2+y^2} \quad \textrm{and} \quad \vec{x}=(x,y,z).
\nonumber
\ee

In some models we also include an additional spiral arm component.
This is taken from Ortiz \& L\'epine \cite{ortiz93}, who obtained a
good fit to the tangent directions seen in infrared star counts with
four logarithmic spirals.  See Table 1 of Englmaier \& Gerhard
\cite{ppe99} for other tracers of these tangent directions. The
positions of the spiral arms $r_i(\phi)$ ($i=1,\ldots,4$) are given by
\be
r_i(\phi)=2.33\kpc \cdot e^{\left(\phi-\phibar-\phi_i \right) 
\cdot \tan(\chi)},
\nonumber\ee
where the angle $\phi_i=0,\pi/2,\pi, 3\pi/2$ determines the starting
angle of a spiral arm in galactocentric coordinates with respect to
the major axis of the bar, and $\chi=13.8\deg$ is the pitch angle of
the arms (Ortiz \& L\'epine \cite{ortiz93}).  We use this 4-armed
logarithmic spiral in the range between an inner radius of $3.5\kpc$
and an outer radius of approximately $10\kpc$. 
We do not ensure a smooth transition to the bar in the parametric 
model. The spiral arms are
modelled with a Gaussian profile with FWHM usually $\approx 300\pc$,
again after Ortiz \& L\'epine, but we have also computed models with
$\approx 500\pc$, without improving the fit to the data as described
below. In our parametric models we treat this spiral arm model as an
enhancement of the density of our standard disc model, keeping all the
above spiral arm parameters fixed and varying only the amplitude $d_{s}$
of the density modulations:
\be
\rho_d^{\rm including\ spiral}=\rho_d\cdot\prod_{i=1}^{4}
\left(1+d_s\cdot e^{-\ln(2)\cdot \Delta r_i^2/(0.5\cdot{\rm FWHM})^2} \right)
\nonumber\ee
where $\Delta r_i$ is the (approximate) distance to the nearest point
on spiral arm $i$. 
The projected density is matched to the {\sl COBE/DIRBE} L-band data
and the best fit parameters are found with our implementation of the
Marquardt-Levenberg algorithm (Press \etal \cite{press}), now with
$d_s$ as an additional fit parameter. These parametric best-fit models
(as a function of the bar angle $\phibar$) are used both as initial
models and to define a spiral arm bias term in the penalty function
(see below) in the non-parametric deprojection of the {\sl COBE/DIRBE}
L-band data.

\begin{figure}
  \getfig{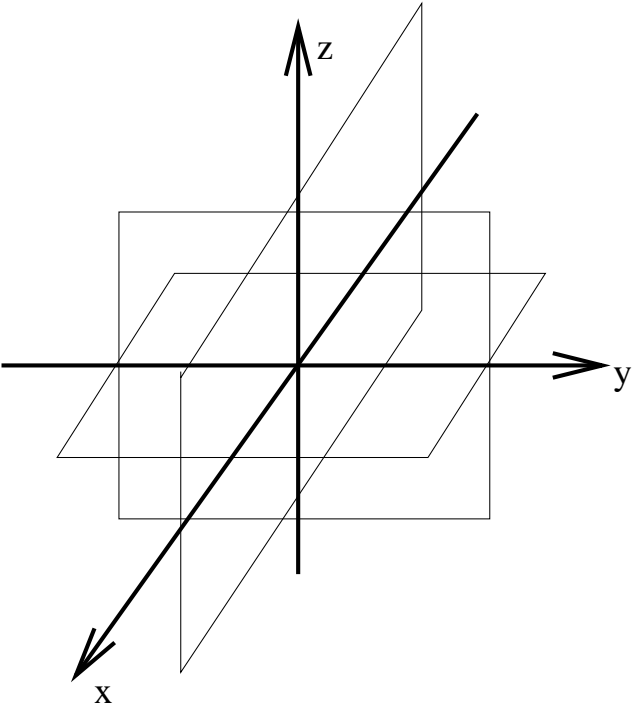}{}
  \caption{The eight-fold symmetry term in the penalty function encourages 
    symmetry of the density function with respect to the mirror planes
    shown in this figure.  Coordinate directions are in the bar frame.
    }
  \label{figoct}
\end{figure}

\subsection{The algorithm}

Our approach is non-parametric: the idea is to maximize
a likelihood function which includes penalty terms encouraging
smoothness, eight-fold symmetry and a spiral structure close to the
imposed four-armed pattern. Thus also the bar by itself is not
forced to obey eight-fold symmetry, but will be nearly triaxially
symmetric as far as allowed by the data and the other constraints.

For the technical realization, the model density is defined on a
cartesian grid. Stepsizes are identical in \(x\) and \(y\). The
stepsize in \(z\) is smaller than for \(x\) and \(y\) because we
expect the most rapid spatial change of the density along \(z\).
The ``standard'' grid consists of $60\!\times\!
60\!\times\! 41$ points covering a box of ${(-5\kpc\!\leq\! x\!  \leq\!
  5\kpc)}\! \times \!{(-5 \kpc\!\leq\! y\! \leq\! 5\kpc)}\! \times\!
{(-1.5\kpc\!  \leq\! z\! \leq\! 1.5\kpc)}$ in $x,y,z$. The size of the
box is chosen so as to emphasize the bar region; outside its boundaries
the parametric model is left unchanged. 
This leads to a discontinuity in the density at the grid boundary; 
for example, averaged over the high-density region $|z|\leq 450\pc$
around the grid boundary, the rms discontinuity is $<1\%$. 
The likelihood maximisation is done using a conjugate gradient method.

The likelihood function ${\cal L}$ maximized by the algorithm is
\begin{eqnarray} \label{eqlikelihood}
L[\ln(\rho)]& =& -(\frac{1}{2}\cdot\chi^2_{\rm SB}+\\
  & & \lambda\cdot D^2[\ln(\rho)]+o\cdot 
  \chi^2_{8} + s\cdot\chi^2_{\rm spiral})
\nonumber\end{eqnarray}
where the individual penalty terms are now described in more detail,
and $\lambda, o, s$ are the penalty parameters. 

\begin{description}
\item[1.Surface brighntess term:]
\begin{equation}\label{sbterm}
\chi^2_{\rm SB}=\sum_{{\rm all}\, {\rm SB}(m,n)} 
\left[\ln\left[{\cal P}(m,n)\right]-{\rm SB}(m,n)\right]^2,
\end{equation}
where ${\cal P}(m,n)$ is the projection of the density $\rho$ along
the {\los} at the sky position $(l_m, b_n)$ of the corresponding {\sl
  COBE/DIRBE} surface brightness data point ${\rm SB}(m,n)$ (natural
log of flux in MJy/sr).  Only the parts of the {\los}
that are in the model grid are taken into account in the projection.
We therefore rescale the observed surface brightness for each {\los} by
multiplying it with the ratio of the surface density in the box to the
total surface density, calculated for the initial parametric model.
${\rm SB}(m,n)$ denotes this box-corrected surface brightness. Outlyers
(data points with a very large distance to the projection of the initial
density $\rho_0$: $\left[\ln\left[{\cal P}(m,n)\right]-{\rm SB}(m,n)
\right]^2\geq 10$) are ignored in the sum eq. (\ref{sbterm}).

\item[2.Smoothness:]
\begin{eqnarray} 
D^2[\ln(\rho)] & = & \sum_{\alpha\beta\in {xx,yy,zz,xy,xz,yz}}
w_{\alpha\beta} \times \nonumber\\
 &  &  \sum_{{\rm Interior}\, {\rm points} (i,j,k)} 
D_{\alpha\beta}^2[\ln(\rho)]\cdot w^{\alpha\beta}_{(i,j,k)}. \nonumber
\end{eqnarray}
This penalty term encourages smoothness of the density distribution by
minimizing the total second derivative.  All partial second
derivatives are taken into account, and are symbolized by their
``coordinate direction'' $\alpha\beta$. For example $\alpha\beta=xy$
stands for $\frac{\partial^2}{\partial x \partial y}$.  All second
derivatives are evaluated only at interior grid points not on any
boundary of the box.  Because the stepsize $s_z$ in the density grid
is smaller than the stepsizes $s_x=s_y$, the six ``coordinate
direction''-terms are given weights $w_{\alpha\beta}$.  These are
$w_{xx}\!\! =\!\! w_{yy}\!\! =\!\! w_{xy}\!\! = 1$, $w_{xz}\!\!=
\!\!w_{yz}\!\!=\!\!\left(\frac{s_x}{s_z}\right)^2$ and
$w_{zz}\!\!=\!\!\left( \frac{s_x}{s_z}\right)^4$.  The functions
$D_{\alpha\beta}$ are first order approximations for the second
derivatives along the ``coordinate direction'' $\alpha\beta$; for
example
\begin{eqnarray}
D_{xx} & =  & 
\ln\left(\rho_{i+1,j,k}\right)-2\cdot\ln\left(\rho_
{i,j,k}\right) \nonumber \\
 &  + & \ln(\left(\rho_{i-1,j,k}\right), \nonumber
\end{eqnarray} 
\begin{eqnarray}
D_{xy} & = & \left[\ln\left(\rho_{i+1,j+1,k}\right)-
\ln\left(\rho_{i-1,j+1,k}\right)\right] \nonumber \\
 & - & \left[\ln\left(\rho_{i+1,j-1,k}\right)-
\ln\left(\rho_{i-1,j-1,k}\right)\right]. \nonumber 
\end{eqnarray}
We expect significant differences in the expected second derivatives
between different parts of the grid. For example high rates of change
of the density are expected at small galactocentric radii.
Therefore we give each grid point and ``coordinate direction'' $\alpha\beta$
additional individual weights  $w^{\alpha\beta}_{(i,j,k)}$.
We have tried two different approaches for these individual  weights.
Using the density $\hat{\rho}$ of the initial non-parametric model,
we have used 
\begin{eqnarray}
w_{(i,j,k)}^{xx} & = & \ln\left(\hat{\rho}\right)_{i,j,k} \nonumber \\
 & / & \frac{1}{4}\cdot\left[
\ln\left(\hat{\rho}\right)_{i+1,j,k}
-\ln\left(\hat{\rho}\right)_{i-1,j,k}\right]^2 \nonumber
\end{eqnarray}
and 
\begin{eqnarray}
w_{(i,j,k)}^{xy} & = & \ln\left(\hat{\rho}\right)_{i,j,k} \nonumber \\
 & / & |\frac{1}{4}\cdot\left[\ln\left(
\hat{\rho}\right)_{i+1,j,k}-
\ln\left(\hat{\rho}\right)_{i-1,j,k}\right] \nonumber \\
 & \cdot & \left[\ln\left(\hat{\rho}\right)_{i,j+1,k}-
\ln\left(\hat{\rho}\right)_{i,j-1,k})\right]|. \nonumber
\end{eqnarray}
or, in the second approach,
$w^{\alpha\beta}_{(i,j,k)}=D_{\alpha\beta}^2
\left[\ln(\hat{\rho}) \right]$. We have found no significant
differences between models based the two approaches.  Therefore we do
not expect a significant influence of the exact definition of these
weights on our results.

We have also tried a smoothing term defined on a cylindrical grid,
using $\rho(r,\phi,z)$ and a smoothness penalty term
\begin{eqnarray} \label{eqD}
D^2[\ln(\rho)] & = & \sum_{\alpha\beta\in {rr,\phi\phi,zz}} w_{\alpha\beta} 
 \sum_{{\rm Interior}\, {\rm points} (i,j,k)} \{\nonumber\\
 &  & 
D_{\alpha\beta}^2[\ln(\rho_c)]\cdot G_{\alpha\beta}(r) 
   \cdot w_{(i,j,k)}^{\alpha\beta} \}.\nonumber
\end{eqnarray}
Here $G_{\alpha\beta}(r)\equiv 1$ for $\alpha\beta=rr \wedge
\alpha\beta=zz$ and $G_{\phi\phi}(r)=r/r_{max}$.  In tests comparing
the two different smoothing penalty terms we have found that the
cartesian smoothing needs somewhat better initial models to give good
final results, while the cylindrical smoothing introduces some bias
towards round models.  However, the main results described in the
sections to follow have been checked by doing the calculations with
both approaches and were found to be identical. Our results therefore
do not depend on the precise smoothing approach and in the following,
the cartesian smoothing will generally be used.
 
To close the discussion of the smoothness penalty term we remark on a
technical detail. The algorithm to maximize the likelihood function
evaluates the gradient $\frac{\partial {\cal L} }{\partial
  (\ln(\rho_{ijk}))}$.  This gradient is modified slightly in the case
of a cartesian smoothness penalty term: terms that couple a point to
its neighbours of second order (i.e., not their nearest neighbours)
are then omitted. We find that without this change the isodensity
contours in the outer parts of the final models become rectangular
because the cartesian smoothness term favours straight contours
parallel to the coordinate axes.

\item[3.Eight-fold symmetry:]
Triaxial 
symmetry 
with 
respect 
to 
three 
principal 
planes 
of
the bar (see Fig.~\ref{figoct}) is an essential requirement for being
able to obtain a three-dimensional luminosity distribution from the
{\sl COBE/DIRBE} surface brightness map (Binney \& Gerhard \cite{bg96}).
Bars in external galaxies are observed to be approximately but not
strictly eight-fold symmetric (e.g., Sellwood \& Wilkinson \cite{sw});
In our deprojection we therefore
aim to find a luminosity distribution that is as nearly eight-fold
symmetric as is compatible with the data and the smoothness constraint.
This is done by discouraging deviations from eight-fold symmetry through
the penalty term
\be
\chi^2_{8} & = & \sum_{i,j,k} \sum_{\rm pairs} 
\left(\ln\left(\rho_
{i',j',k'}\right)-\ln\left(\rho_{i'',j'',k''}\right)\right)^2.
\nonumber\ee 
Here the inner sum is taken over all distinct pairs of grid points
constructed from the eight mirror-symmetric points of grid point
$(i,j,k)$, that should have identical luminosity
density if the distribution were fully eight-fold symmetric.

\item[4.Spiral structure term:] 

Generally there is not enough
  information in the {\sl COBE/DIRBE} surface brightness data to determine
  the luminosity distribution in the Galactic spiral arms. Essentially
  the only information about the spiral arms in these data is an
  enhanced surface brightness in the arm tangent directions (see also
  Drimmel \& Spergel \cite{drimmsp01}).  Deriving a sensible model
  therefore requires using additional, external information on the arm
  pattern.  For most of our models we assume that the Galactic spiral
  arms in the NIR are described approximately by the four-armed
  pattern that seems to be most consistent with observations of HII
  regions, gas, young stars, and NIR starcounts (see Ortiz \& L\'epine
  \cite{ortiz93} and the summary in Englmaier \& Gerhard
  \cite{ppe99}). We leave open the question whether the old
  population of the Galactic disk follows this four-armed or rather a
  two-armed pattern.  In practice, we discourage deviations from the
  disk part of the initial parametric model $\hat{\rho}$ that is also
  used to start the iterations, which includes the Ortiz-L\'epine
  spiral arm model (Section \ref{secparametric}).  The penalty term
  is:
\begin{eqnarray}\label{eqspiralterm}
\chi^2_{\rm spiral} & = &  \sum_{k:\ |z(k)|\leq z_s} \sum_{i,j}
  \left[w(k) \right. \\
 & + &  \left. \ln(\hat{\rho}_{i,j,k})-
        \ln(\rho_{i,j,k})\right]^2 \nonumber
\end{eqnarray}
with
\be
w(k)  & =  & \ln\left(\frac{\sum_{k,{\rm all}\, i,j} 
  \rho_{i,j,k}}{\sum_{k,{\rm all}\, i,j} 
  \hat{\rho}_{i,j,k}}\right).
\nonumber\ee
The outer sum in eq. (\ref{eqspiralterm}) is computed over all planes
parallel to the main plane of the MW with indices $k$ for which
$|z(k)|\!\leq\! z_s$ with $z_s \! = \! 300\pc$, and the inner
sum is taken over all points within the current plane.  I.e., the
model is only biased towards the initial model near the disk plane. 
The weights $w(k)$ guarantee that the
model is encouraged to resemble $\hat{\rho}$ only {\sl in shape}, but
not in normalisation.  In fact, the normalisation ratios $w(k)$
are usually somewhat different for every iteration step.
We have tested that restricting  the spiral structure penalty 
term to $R\geq 3.5\kpc$ did not change our models significantly.
\end{description}

\subsection{Optimal penalty parameters from test models}

Having specified the likelihood function (\ref{eqlikelihood}), we now
need to determine the penalty parameters $\lambda, o, s$ which set
the relative importance of the different penalty function terms.
These can be found approximately using known test models by requiring
that all terms in the penalty function should be of the same order of
magnitude. Otherwise one of the imposed constraints would be given too
high or too low weight in the resulting model. Because
non-parametric models do differ from the test models employed in this
determination, we found it necessary to vary the resulting penalty
function parameters within an order of magnitude or so, based on a
(subjective) by-eye assessment of the final model.

The special properties of the spiral arm bias term require an
additional modification of this simple scheme. The spiral structure is
confined to the vicinity of the Galactic plane, and there it is
neither very smooth on the scale of a few grid cells nor is it
eight-fold symmetric.  On the other hand, it penalizes deviations from
some model $\hat{\rho}$, which has a similar effect like a
regularization term. After some experimentation we found that a good
solution is to change the eight-fold symmetry parameter to $20\%$ of
its original value in the main plane of the MW and let it linearly
rise with distance $|z|$ from the plane to its overall value.
Similarly the smoothness penalty parameter is set to only $1$ percent
of its maximum value in the main plane, but rises very fast $\propto
(1-(1-|z|/1.5\kpc)^{10})$ with distance from the central plane.

For part of the problem (data, smoothness, and eight-fold symmetry) we
have tested the above choice of penalty parameters with a second
approach.  We selected two parametric models without spiral structure,
that differed by about the same amount that we expect our initial
models in deprojections of the {\sl COBE/DIRBE} L-band data to differ
from the ``true'' model. From one of these models we generated
artificial data by projecting it onto the sky and adding Gaussian
noise with $\sigma= 0.076^{\rm mag}$. This is the remaining rms NIR
{\sl colour} variation found by Spergel \etal \cite{spergel96} for
this dereddened {\sl COBE/DIRBE} L-band data set.  We then deprojected
these data using the second parametric model as initial model and
repeated the deprojections for a grid of points in $(\lambda,o)$
space. For all deprojected models we computed the rms difference
between data and projection of the model onto the sky, and the rms
difference in the natural logarithm between the ``true'' model and the
deprojected model on the density grid. For computational reasons
we could do this only for a coarser density grid. The resulting
optimal penalty parameters $\lambda$, $o$, when rescaled to the
original grid, agree with the values obtained by the ``equal penalty
terms'' method to within 1-2 orders of magnitude.

\subsection{Performance and limitation of the algorithm}
\label{sectperform}

We have tested this algorithm with noisy artificial data and initial
conditions derived from a variety of test models.  Quantitative
results will be given in the next section, which studies ambiguituies
in the deprojection of Galactic bar and disk models from surface
brightness data, under the assumptions made. Here we discuss only a few
qualitative points.

It is clear that information about the ``true'' model used to generate
the artificial data is increasingly hard to recover as the noise level
approaches the magnitude of the signal that differentiates between
different models on the sky. On the other hand, we have found that
some noise is helpful as a ``catalyst'' to induce changes in the
model.

The initial models given to the algorithm differed from the ``true''
model by various amounts. We find that the initial model must not be
too far from the ``true'' model. This is hard to quantify by a
distance criterion.  However, the effects that occur if the initial
model is not suitable are easily visible 
 in cuts through the density grid, and it is therefore
possible to reject such models.

If the initial model is suitable, convergence to a luminosity
distribution that fits the surface brightness map under the assumed
smoothness, symmetry and spiral structure constraints typically takes
20-30 iterations. Otherwise the progress of the iterations becomes
very slow at some point and the model may be caught near something
like a secondary minimum. In such cases, an iteration step that would
be required to improve the surface brightness fit is often not
undertaken because it would move the model too far away from
eight-fold symmetry. Some secondary minima of the likelihood function
correspond to nearly perfectly eight-fold symmetric models which have,
however, physically unreasonable density distributions. The
probability that the algorithm ends up with such a model increases 
with the distance between the initial model and the ``true'' model.

Introducing a smoothness term in a complicated $\chi^2$ fitting
problem often lessens the importance of such secondary minima. 
Here the additional problem is that a third constraint,
eight-fold symmetry, must be introduced to restrict the range of
possible solutions (for fixed bar angle, the requirement of eight-fold
symmetry restricts the solution to a small subset of the very large
set of luminosity distributions which all project to the same surface
brightness distribution, Binney \& Gerhard \cite{bg96}). It is easy
to see then that secondary minima based on a balance between data and
symmetry terms can appear in spite of the smoothing.

If the requirement of eight-fold symmetry is imposed only weakly, a
characteristic artifact appears in many models which we have termed
{\sl Finger-to-Sun} (FTS) effect. This consists of excess luminosity
features in the nearby disk pointing towards the observer. These arise
because the deprojection algorithm preferentially changes grid cells
near the observer. The reason for this is that grid cells near the
observer appear larger on the sky, and therefore contribute to many
more surface brightness pixels than distant grid cells do.
Consequently, for a model that is off from the ``true'' model
underlying the data by a fixed fraction of the density in all grid
cells, the total gradient ${\partial \chi^2_{\rm SB}}/{\partial
  \left(\ln\rho_{ijk}\right)}$ is much larger for grid cells near the
observer (the actual value contributed to the surface brightness of a
given pixel is independent of the line-of-sight distance of the
contributing grid cells).  Therefore, without smoothing and symmetry
penalty terms, the luminosity model would be changed mainly near
the observer, resulting in the described FTS effects.

\section{How well-determined is the deprojection of
the bar?}
\label{secseq}

\begin{figure*}
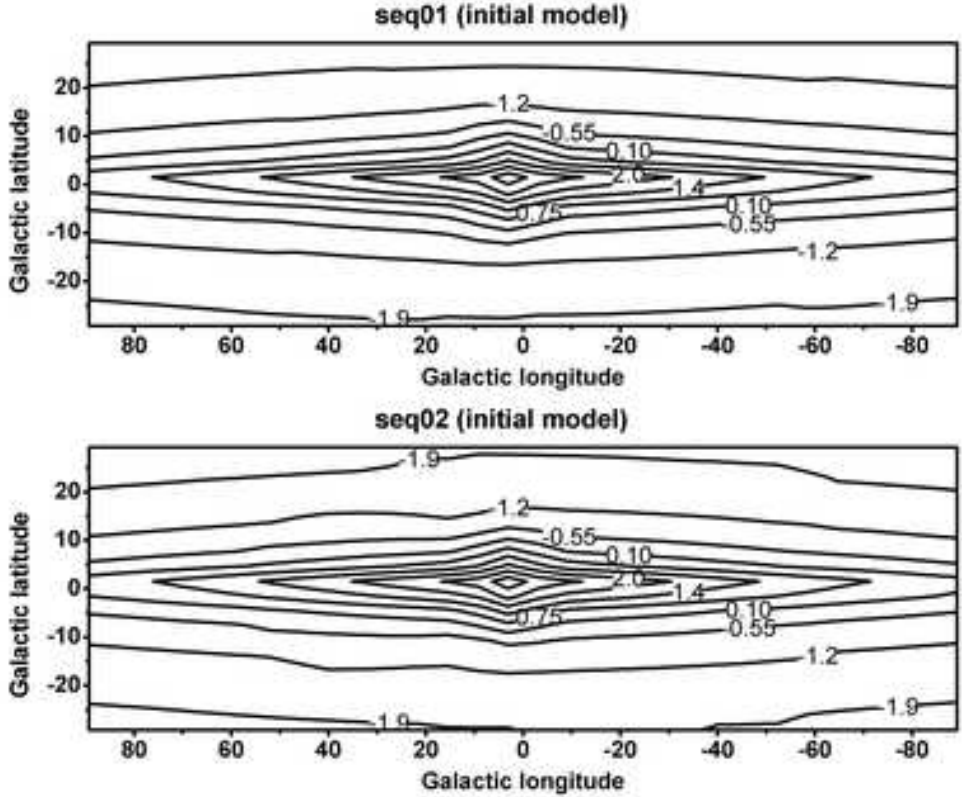

  \getfig{fignusb.eps2}{}
  \caption{Two parametric models on the degenerate sequence
for $\phibar=20\deg$. Note how similar models s1p (upper panel) 
and s2p (lower panel) appear to the observer.}
  \label{fignusb}
\end{figure*}    

\begin{figure}
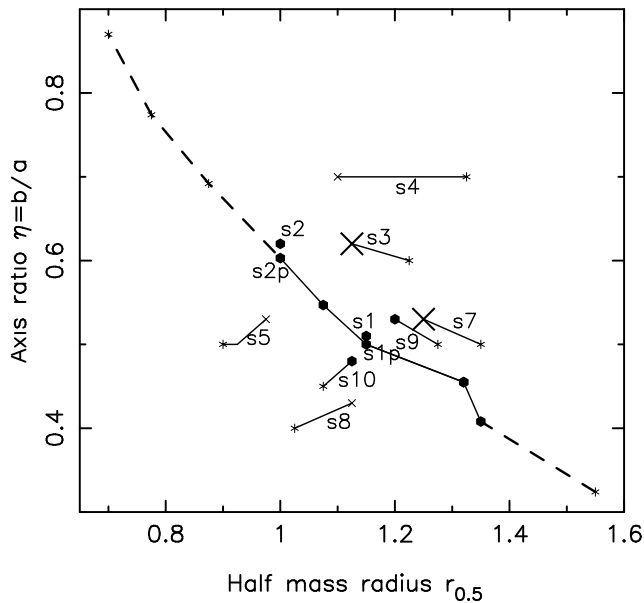

  \getfig{sequenzplot.eps2}{}
  \caption{Parametric and non-parametric luminosity models for
    surface brightness data obtained from projecting model s1p under
    $\phibar=20\deg$ and adding noise. 
    All models are plotted in a plane according to
    their half-mass radius $r_{0.5}$ and bar elongation $\eta$.  The
    full line delineates the sequence of degenerate parametric models
    which project to indistinguishable SB data. These models are
    indicated by the filled hexagons along this line. The dashed lines
    connect parametric models (stars) on the extensions of this
    sequence towards high and low $\eta$. These models are not
    acceptable, because they either show too large systematic
    devations from the data, or have quality grade $F>3$ (see
    eq.~\ref{eqqualityF}).  Models s1-s10 are obtained from
    non-parametric, iterative deprojections of the model SB data. Of
    these, s1,s2 were started from parametric models (s1p, s2p)
    along the sequence.  Their proximity to the original (s1p, s2p)
    demonstrates the absence of significant bias in the algorithm.
    The other non-parametric models were started off the sequence
    (stars on one end of the short lines denote the initial
    configurations) and are separated in the figure into acceptable
    final models (filled hexagons, $F\leq3$), marginally acceptable
    models ($3<F\leq4$, large ``X''), and inacceptable models ($F>4$,
    small ``x'').  }
  \label{figsequence}
\end{figure}    

\begin{figure}
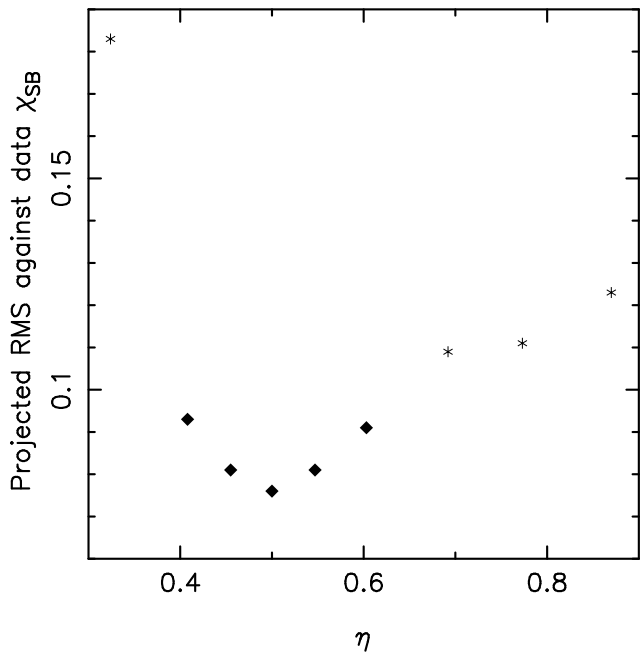

  \getfig{figpararms.eps2}{}
  \caption{RMS SB residuals, $\chi_{\rm SB}$, in magnitudes, for 
    parametric models on the degenerate sequence, plotted as a
    function of in-plane axial ratio $\eta$ (equivalent to varying
    $r_{0.5}$). Diamonds indicate models on the sequence, stars models
    on the extensions of the sequence.}
  \label{figpararms}
\end{figure}

\begin{figure}
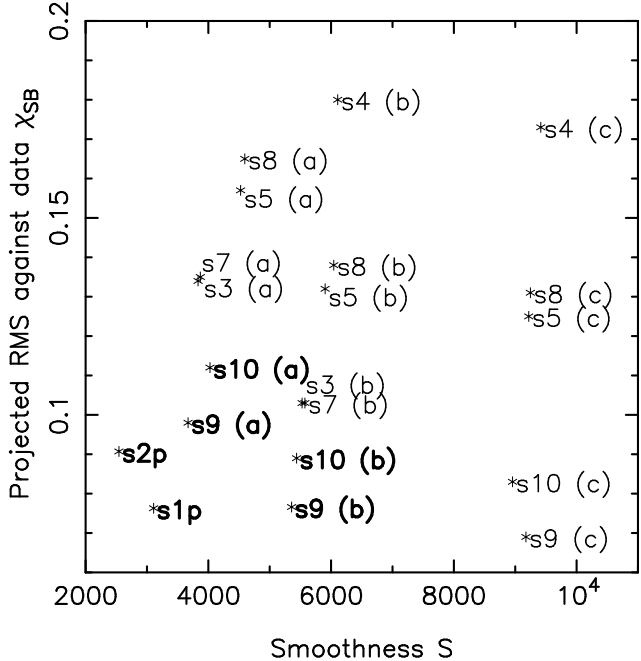

  \getfig{fig_chiintrplot.eps2}{}
  \caption{Non-parametric luminosity models s3-s10 obtained from model
        SB data with the algorithm described in Section
    \ref{secmldmethod}. The parametric models s1p from which the data
    were generated, and s2p on the degenerate sequence, are
    also shown.  The models are characterized by their
    smoothness $S$ (smaller $S$ means smoother), and by their rms
    difference $\chi_{\rm SB}$ with respect to the SB data on the sky, in
    magnitudes.  Most models have been calculated with three values of
    the parameter $\lambda$; the last symbol ``a'',''b'' or ``c'' in
    the non-parametric model name (for example s4a) indicates
    $\lambda$, decreasing from ``a'' to ``c'' in multiplicative steps
    of $10$. Model names printed in bold denote acceptable models,
    i.e., have  quality grade $F\leq3$ (see eq.~\ref{eqqualityF}). }
  \label{fig_chiintrplot}
\end{figure}

As a test application of our algorithm, we investigate in this section
possible degeneracies in the deprojection of a bar model for fixed bar
angle, keeping $\phibar=20\deg$ throughout this section.  
More precisely, we ask what is the
range of bar luminosity distributions that is compatible with given
surface brightness data similar in quality to the {\sl COBE} data which we
will use in \S5 for investigating the luminosity structure of the
inner Galaxy. We find that there exists a sequence of parametric
models with different bar elongations in the Galactic plane and
correspondingly different density concentrations, that look very
similar on the sky. Then we use the non-parametric algorithm to
estimate the width of the ``valley'' of acceptable models around the
sequence of degenerate parametric models. Finally, we show that
observations such as the apparent magnitude distributions of clump
giant stars by Stanek \etal \cite{stanek97} contain sufficient distance
information to break this degeneracy, which incidentally is different
from the well-known degeneracy in the bar angle $\phibar$ (Binney,
Gerhard \& Spergel \cite{bgs97}, Zhao \cite{zhao00}).

We first generate artificial data by projecting a parametric model
onto the sky. This model (denoted s1p) is defined by eq. (\ref{eqpara}), 
with bar parameters $\eta=0.5, \zeta=0.6, a_m=1.8$ and
disk parameters $z_0=208\pc, R_d=2.44\kpc$, and does not have spiral
arms.  We add Gaussian noise to these data, similar in amplitude to
that expected for the {\sl COBE/DIRBE} L-band data used later, for
which $\sigma_{\rm SB}=0.076^m$ (Spergel \etal \cite{spergel96}, see
Section \ref{sectnospiral} for a discussion). The parametric model is a
best-fit model for the {\sl COBE/DIRBE} L-band data for $\phibar\!=\!
20\deg$. 

Next we fit other parametric models to these model surface brightness
data.  In these models we hold fixed the bar elongation $\eta$. To
decide whether a model is a valid match to the data, we define two
criteria: (i) The average square deviations of the projected model
from the artificial data in magnitudes (hereafter model RMS) must not
be worse by more than $20\%$ in the bar region $|l|\! <\! 36\deg
\wedge |b|\! <\! 11\deg$ than for the ``true'' model. (ii) There must
not be major parts of this region where a systematic error larger
than the approximate average statistical error occurs in the residual
map.

With these criteria, we find a sequence of parametric models that are
indistuingishible on the sky, i.e., for which systematic deviations
between model and data are smaller than the noise in the data. Figure
\ref{fignusb} compares model s1p with another parametric model
s2p, which is on this sequence.  Model s2p has bar parameters
$\eta=0.603, \zeta=0.68, a_m=1.5\kpc$, significantly different from
model s1p, but looks very similar on the sky.  Models on this
sequence are characterized by a degeneracy between the input bar
elongation in the $xy$-plane, $\eta$ (see eq.~\ref{eqpara}), and the
central concentration of the model. This is parametrized as the half
mass radius $r_{0.5}$, defined as the elliptical radius
$r_{0.5}=\sqrt{x^2+\frac{y^2}{\eta^2} +\frac{z^2}{\zeta^2}}$ which
contains half of the mass of the bar/bulge inside an elliptical radius
of $3.5\kpc$.
 
In Fig. \ref{figsequence} the sequence of parametric models is
depicted in the $r_{0.5}-\eta$ plane as the filled hexagons connected
by the full line.  The parametric model at the lower-right end of this
sequence is just still a valid model as defined above. For models with
even smaller $\eta$ the deviations from the artificial data rise
rapidly. Fig.~\ref{figpararms} shows the model RMS on the sky, for
parametric models both on the sequence and on its extensions to higher
and lower $\eta$, where the latter fail to pass our criterion for a
valid model.  On the high-$\eta$ extension of the sequence, the model
RMS rises fairly slowly, but these models have regions with too large
systematic deviations from the projected data, which increase with
increasing $\eta$.  From Fig.~\ref{figsequence} parametric models thus
have an uncertainty in their model parameters of $\pm 0.1$ in $\eta$
and $\pm 20\%$ in $r_{0.5}$ for given data on the sky.

We now consider non-parametric models obtained with the deprojection
algorithm described in \S\ref{secmldmethod}. First, we start the
algorithm from initial models on the parametric sequence, resulting in
models s1 and s2 from the initial s1p and s2p.  For the
non-parametric models we estimate $\eta$ by measuring the elongations of
the surface density contours of the bar, determined from 
a projection of the
model density along the $z$-axis, but excluding the region $|z|\leq
225\pc$ to reduce the impact of the strong axisymmetric disk.  
This is a good approximation because the bulges in these models
are near-ellipsoidal. We
measure the half-mass radius $r_{0.5}$ from the density distribution
of the bar/bulge only, obtained by subtracting the disk density of the
parametric initial model given to the algorithm, from the density
distribution of the final non-parametric model. We have checked by
visual inspection that the resulting bulge densities are reasonable.
Fig.~\ref{figsequence} shows that models s1p and s2p lie very
close to s1 and s2. The similarity of initial and final models is
also obvious from a comparison of cuts through the densities.  We
conclude from this exercise that the deprojection algorithm does not
introduce any significant bias in the final model, e.g., in the
resulting value of $\eta$.

We will next discuss non-parametric models started from parametric
models off the degenerate sequence, in order to investigate how broad
the valley of acceptable models surrounding the sequence is.  Several
such models (s3-s10) are shown in the $r_{0.5}-\eta$-plane of
Fig.~\ref{figsequence}, as stars for the respective initial parametric
models, and as hexagons and crosses for the final non-parametric
density distributions after around 50 iterations.  Clearly the algorithm
evolves these models to the vicinity of the parametric sequence.
Whether these non-parametric models are acceptable cannot be decided
only on the basis of the model RMS on the sky, however.
For in a non-parametric model, substantial grid cell to grid cell
noise can be introduced in order to improve the match to the data,
which beyond a certain point is clearly unphysical.  Therefore some
measure of smoothness must be introduced in judging a model's
validity.

We measure the smoothness $S$ of some model M as the sum of the
absolute differences in logarithmic density, between M and a smoothed
version of M. S will be small for smooth M and large for noisy M. In
determining S we sum only over grid cells with $|z|\leq 750\pc$, to
avoid contributions from fluctuations in regions of the density grid
where the density is very small.  We smooth a model as follows,
working with logarithmic model densities: (i) We resample each
$z$-plane of the model on a cylindrical grid of $30$ linearly spaced
points in $r=\sqrt{x^2+y^2}$ out to $r=7\kpc$, and $60$ points in
azimuth $\phi$. (ii) We smooth the model over $5$ points, first in the
$\phi$-coordinate, then in $r$, and finally in $z$, using second order
polynomials (Savitzky-Golay filters, see Press \etal \cite{press}). In
this way azimuthal gradients in the central parts of the model are not
smoothed away. (iii) We re-interpolate to the original cartesian grid.
(iv) In models with spiral arms, this procedure must be modified
because the spiral arms imply density changes on small scales.  For
example, on a circle at galactocentric radius $5\kpc$, the distance
between adjacent points in $\phi$ in our smoothing algorithm is
$\approx 500\pc$; therefore the smoothing length is of the same order
as the spiral arm FWHM. Thus before actually smoothing the model, we
subtract from the density at every grid point the density of the
spiral arm contribution in the initial parametric model, rescaled in
each $xy$-plane separately. The rescaling factor for each $xy$-plane
is determined by requiring that in this plane the rescaled mass of the
initial model is the same as that of the (non-parametric) model that
we actually smooth.  Having subtracted the rescaled spiral model, we
smooth the remaining luminosity distribution as in (ii), and then add
the subtracted spiral density back to the smoothed density. This
spiral arm preservation procedure is restricted to model planes with
$|z|\leq450\pc$, because the initial spiral models only extend to this
height. The procedure ensures that the spiral arm component does not
contribute to the final difference between M and the smoothed version
of M, i.e., to S.

\begin{figure*}
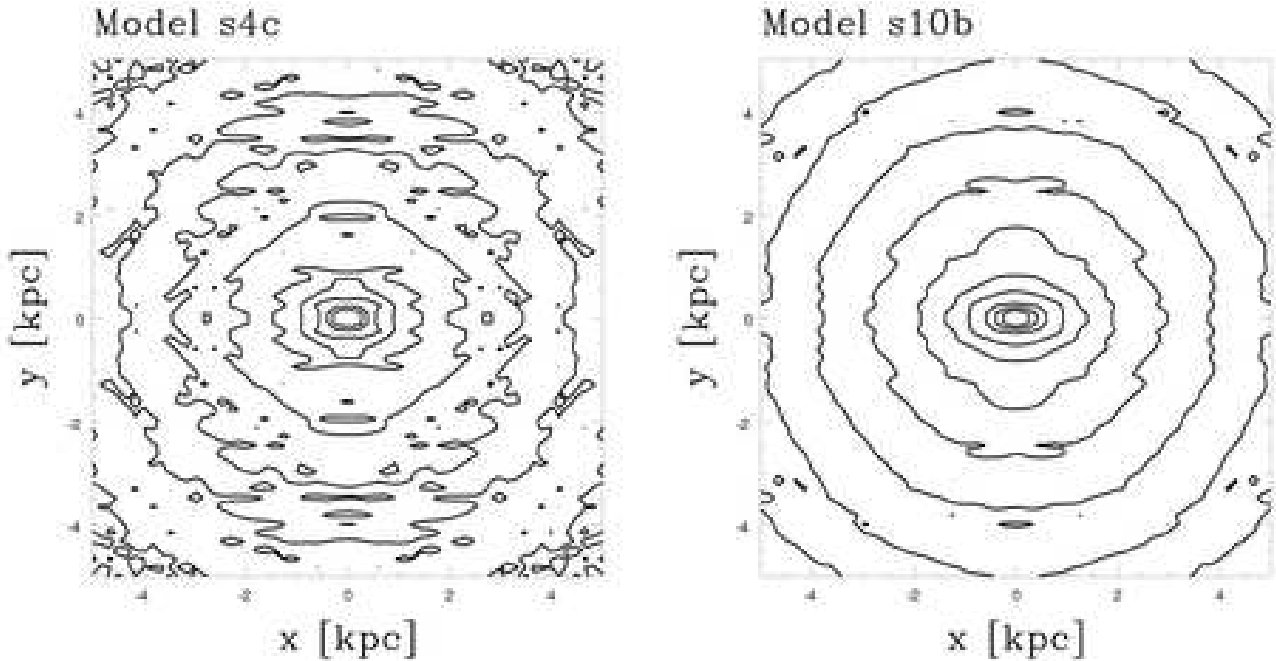

\getfig{fig_smoothexample.eps2}{}
  \caption{A comparison of cuts through two models with significantly 
different smoothness $S$ at $z\!=\!75\pc$. Left: Model s4c 
($S\approx 9400$) Right: Model s10b (with $S\approx 5400$). 
We consider model s4c not smooth enough to be acceptable.}
  \label{fig_smoothexample}
\end{figure*}

Figure \ref{fig_chiintrplot} shows the final
non-parametric models considered in this section in a plane of model
RMS $\chi_{\rm SB}$ and smoothness $S$. The best non-parametric models
have $S\simeq 4000$, parametric models have typically $S\simeq 3000$,
and models with $S\gta 7000$ are not smooth enough to be acceptable.
We illustrate this in Figure \ref{fig_smoothexample}, which shows cuts
through model s4c (which has $S\approx 9400$) at $z=75\pc$ and model
s10b (with $S\approx 5400$), also at $z=75\pc$.

We can now define a criterion to decide whether a non-parametric model
is an acceptable representation of the SB data.  To this end we introduce
the quality grade
\begin{equation}\label{eqqualityF}
F=\left(\frac{\chi_{\rm SB}}{0.076^m}\right)^2 + \left(\frac{S}{S_0}\right)^2,
\end{equation}
where $S_0$ is a measure of $S$ for the ``best'' non-parametric models
we find.  We use the average $S_0=4827$ of the three non-parametric
models s9(a), s9(b), s10(a) printed in bold on Fig.~\ref{fig_chiintrplot}.
$F$ is smaller for models that fit the data better and that are smoother.
The maximum value of $F$ for an acceptable model is somewhat subjective;
we decided for $F\leq3$.  
This results in $\chi_{\rm SB}\lta 0.11^m$ for a model with $S\sim S_0$
and $S\lta 6800$ for a model with $\chi_{\rm SB}\sim 0.076^m$. 
Models which 
violate the last criterion are not smooth enough to be viable 
(see Figure \ref{fig_smoothexample}).

We have used the $F$-criterium to separate the final non-parametric
models s3-s10 in Fig.~\ref{figsequence} into valid and non-valid
models. Filled hexagons indicate acceptable models ($F\leq3$),
large ``X'' marginally acceptable models with $3<F\leq4$, and small
``x'' unacceptable models with $F>4$ (of the three models computed
with different $\lambda$, that with lowest $F$ is always shown).
The hexagons thus mark the width of the valley of acceptable models
associated with the sequence.  This width, set by the model RMS and
smoothness, translates to an uncertainty in the structural parameters
of a non-parametric model of $\approx \pm 0.1 \kpc$ in the half-mass
radius $r_{0.5}$ and $\approx\pm0.05$ in the bar elongation $\eta$.
These are smaller than the uncertainties due to the existence of the
degenerate sequence itself.

\begin{figure*}
  \getfig{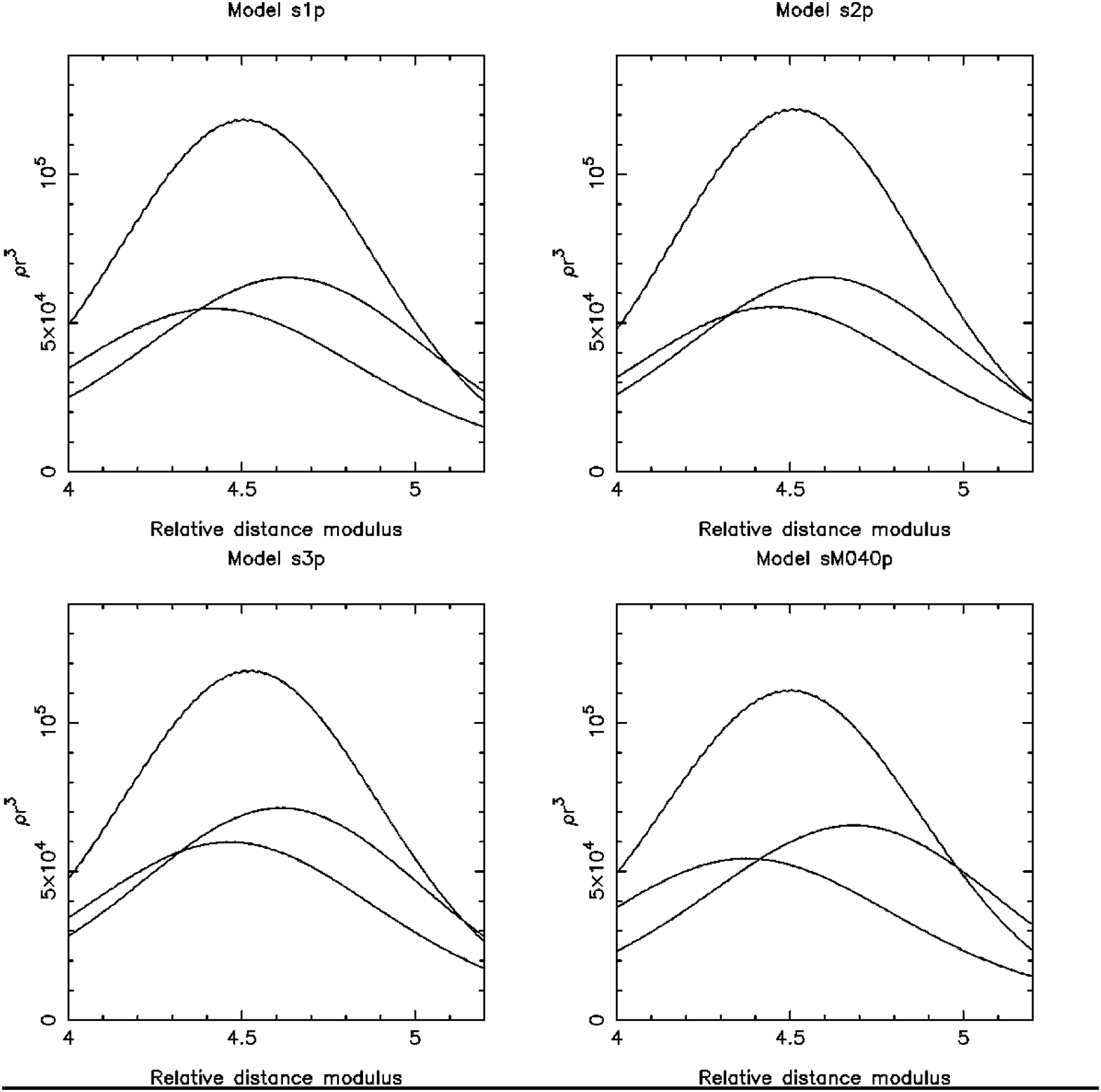}{}
  \caption{Predicted number of clump giant stars in three fields observed
    by Stanek \etal (1997), versus relative distance modulus in magnitudes, for
    parametric models s1p (upper left), s2p (upper right), s3p
    (lower left), and the parametric model with bar elongation $0.41$
    also along the degenerate sequence in Fig.~\ref{figsequence}
    (lower right).  From the differences in height and location of the
    peaks in these distributions, it is possible to discriminate
    between models on the sequence.  However, a comparison of the two
    panels on the left shows that some models off the sequence can
    mimic the clump giant distributions of models on the sequence.
    Such models have to be discriminated by their goodness-of-fit
    for the surface brightness data.}
  \label{fignuclump}
\end{figure*}    

Having quantified the uncertainties due to the extent and width of the
degenerate sequence, we now show that these uncertainties can be much
reduced if addtional distance information is used, such as is
available in the apparent magnitude distributions of clump giant
stars.  Stanek \etal (\cite{stanek94}, \cite{stanek97}) have observed clump giant
distributions in a number of fields towards the bulge. In
Fig.~\ref{fignuclump} we plot such distributions for three models on
the parametric sequence and one at the edge of the
valley of acceptable models around it.  The relevant quantity is $\rho
r^3$, which has one extra power of $r$ over that from the volume
effect, due to the conversion of distance to magnitudes.  The
predicted distributions are different for the models on the sequence.
Depending on the errors in the observations, it is thus possible to
discriminate between these models through their clump giant
distributions.  The clump giant distributions of models 
s3p and s1p {\sl are} very similar; however, these models can be
discriminated on the basis of their projected surface brightness maps,
using e.g., the $F$-criterium. Model s3p is not an acceptable
model for the data generated from s1p. The experiments conducted
in this section make it likely that using 
the goodness-of-fit $F$ and the clump
giant constraints together breaks the degeneracy in bar/bulge models
that exists for fixed $\phibar$.

\section{Deprojection of the inner Galaxy}
\label{secdeprocobe}

In this section we apply the algorithm to the {\sl COBE/DIRBE} data,
using the dust-corrected L-band map of Spergel, Malhotra \& Blitz
\cite{spergel96}. These authors took the $240\mu m$ {\sl COBE/DIRBE}
map as a tracer for the distribution of dust to correct the {\sl
  COBE/DIRBE} near infrared (NIR) J,K,L and M-band data for dust
absorption.  They simultaneously fitted parametric models for the dust
and stars and with these models computed dereddened NIR surface
brightness maps.  The K-band emission near $\lambda\!\approx\! 2.2\mu
m$ is dominated by starlight and only moderately affected by dust. In
the L-band, near $\lambda\!\approx\! 3.5\mu m$, emission by hot dust
and interstellar gas may be slightly more important, but dust
extinction is reduced by about a factor of two in magnitudes as
compared to the K-band. Because in some inner Galaxy regions
extinction is significant even in the K-band, we have decided to use
the L-band data in this paper.

After dereddening Spergel \etal \cite{spergel96} found a mean
dispersion $\sigma\approx 0.076^m$ in colour between the K- and L-band
maps. If we could assume identical Gaussian noise in both maps,
$0.076^m/\,\sqrt[]{2}$ would be a straightforward value to use for the
SB error in the L-band, $\sigma_{\rm SB}$.  However, the dominant sources
of noise are probably systematic errors in the dust correction,
correlated over several pixels and between the NIR maps, especially
near the galactic equator.  In this case the true errors in the data
would be larger. We therefore take a more conservative approach and
use $\sigma_{\rm SB}\approx 0.076^m$. In the non-parametric deprojections
of the NIR data described below we have therefore tailored the
smoothness penalty function parameters such that we get models with
RMS of this order.

The models that we obtain from deprojecting the 
{\sl COBE/DIRBE} L-band data
will then be verified a posteriori by comparing with the apparent
magnitude distributions for clump giant stars, measured by Stanek
\etal (\cite{stanek94}, \cite{stanek97}) along
certain lines-of-sight towards the Galactic bulge.  Clump
giants have nearly identical absolute magnitudes with a small
dispersion of $\sim $0.2-0.3$^m$, and it is therefore possible to
derive their distance distribution (in a statitistical sense) from
their observed magnitude distribution.  Stanek
\etal \cite{stanek97} analysed colour magnitude
diagrams (CMDs) in several OGLE fields, including Baade's window and
two nearly symmetric fields at $(l,b)\!\approx\! (-4.9\deg,-3.4\deg)$
and $(5.5\deg,-3.4\deg)$.  They determined extinction-insensitive
magnitudes $V_{V-I}\!=\!V\!-\!2.6\!\cdot\!  (V\!-\!I)$, and plotted
histograms of the number of stars as a function of magnitude using
$\Delta V_{V-I}=0.05^m$-bins for the stars in the part of the CMD
dominated by bulge red clump stars.  The red clump distributions along
these {\los} peak at different distances; using a bootstrap technique
Stanek \etal \cite{stanek94}  determined a relative distance modulus 
of $0.37\pm 0.025^m$
between the {\los} at $l\!\approx\! 5\deg$ and $l\!\approx\! -5\deg$,
and $0.15\pm 0.02^m$ between Baade's window and the field at
$l\!=\!-4.9\deg$.  These asymmetries provide independent evidence for
a non-axisymmetric luminosity distribution in the inner few $\kpc$ of
the MW, but will be used here to check the three-dimensional
luminosity distribution of our models for the L-band flux data.

In \S\ref{sectnospiral} and \S\ref{sectbarmod} we deproject the data
with and without inclusion of spiral structure in the model, using a
bar angle $\phibar\! =\!  20\deg$ in both cases.  This will
demonstrate that inclusion of spiral structure leads to a better model
for the {\sl COBE/DIRBE} L-band data and results in a more elongated
shape for the barred bulge.  In \S\ref{barangle} we constrain the
acceptable range of bar angle from a set of models with different
$\phibar$ together with the clump giant data.

In all cases, the non-parametric density estimation procedure begins
with fitting a parametric model to the data.  This is used in the
non-parametric deprojection in three ways: (i) As starting model of
the iterations, (ii) to correct
for the limited size of the model density grid and
(iii) in models that include spiral structure, to define the spiral
structure penalty function term.

\begin{figure*}
  \getfig{montage.eps2}{}
  \caption{Surface brightness maps of the model without spiral arms (upper
    panel), and our reference model 20A including spiral arms (lower
    panel). Full contours show the model surface brightness, dashed
    contours the {\sl COBE/DIRBE} L-band data. Contour levels are in
    magnitudes with some arbitrary offset, common to both panels.
    Both surface brightness maps are very similar and fit the {\sl
      COBE/DIRBE} L-band data with very similar $\chi^2$.  The
    underlying models are non-parametric on a grid of 5x5x1.5 kpc$^3$
    and are continued by the initial parametric models outside of this
    grid for the projection onto the sky.}
  \label{figsbmaps}
\end{figure*}

\begin{figure*}
  \getfig{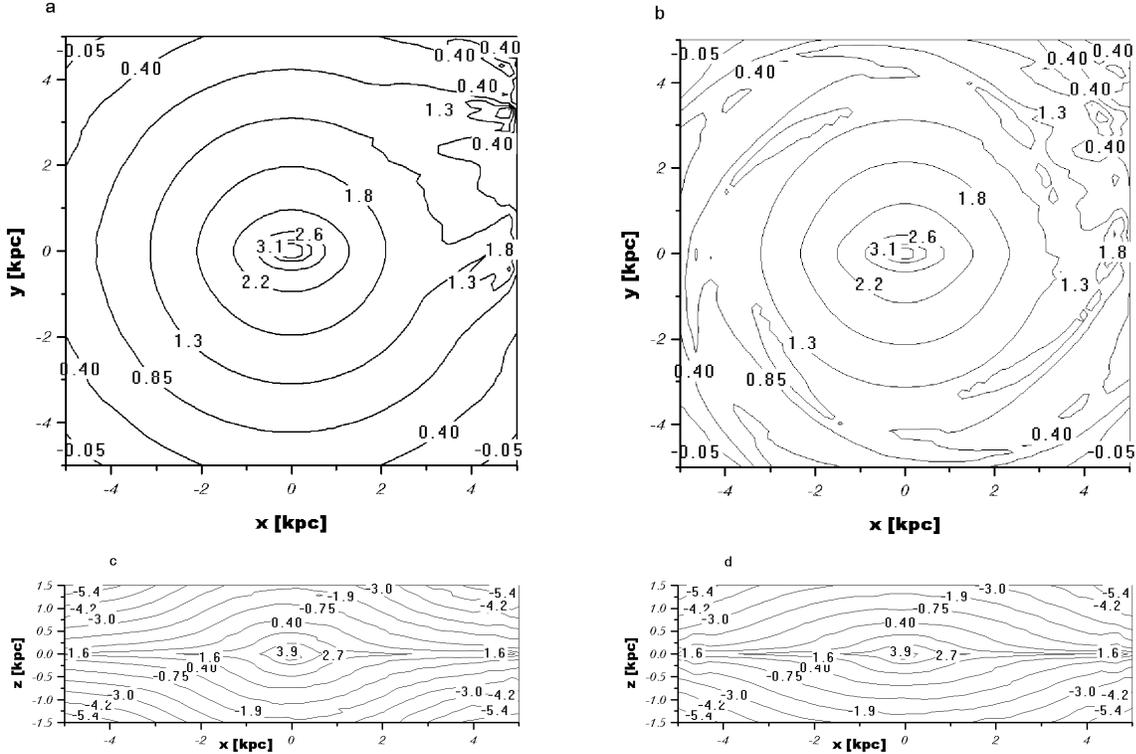}{}
  \caption{Comparison of the models with and 
    without spiral structure: in the main plane of the MW (upper
    panels), and in a plane parallel to the major and minor axes of
    the bar at the first grid point 
    $y\!\approx\! 85\pc$ (lower panels). Both models are
    for bar angle $20\deg$.  Contours are in logarithmic density (in
    $CLU/\kpc^3$), and
    printed in the plot.  Left: The model without spiral structure,
    obtained from a (parametric) triaxially symmetric initial bar
    model. This model shows deviations from eight-fold symmetry in the
    $xy$-map. These overdensities approximately point from the
    observer (at $x\!\approx\! 7.5\kpc$ and $y\!\approx\! 2.7\kpc$)
    towards the tangential points of the spiral arms. Right: Our
    reference model 20A including spiral arms. Both the (parametric)
    initial model and the penalty function in this non-parametric
    density estimation contain a spiral structure term.}
\label{figmodel}
\end{figure*}

\begin{figure}
\getfig{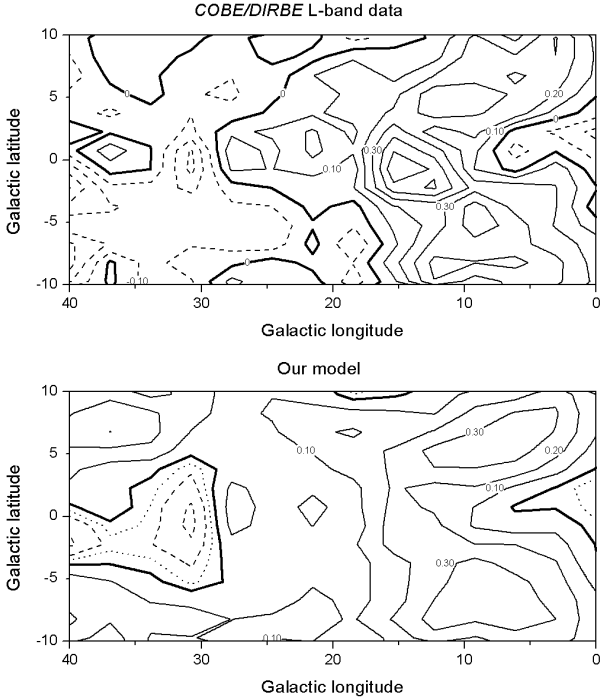}{}
\caption{Asymmetry maps for the {\sl COBE/DIRBE} L-band data
  and for the reference spiral model. The {\sl COBE} data were smoothed
  with a Savitzky-Golay filter (Press \etal 1994, setting their $M=5$) 
  to reduce noise.  Contours show the asymmetry in magnitudes between
  positive and negative longitudes. 
  Dashed contours indicate negative values. 
  Positive values indicate that the
  MW is brighter at positive longitudes. Contour spacing is $0.1^m$,
  and the bold contour is at $0^{m}$ (no left-right asymmetry).
  In the plot for the model
  a dotted contour is drawn additionally at $-0.05^m$.}
\label{figasymmap}
\end{figure}

\begin{figure}
  \getfig{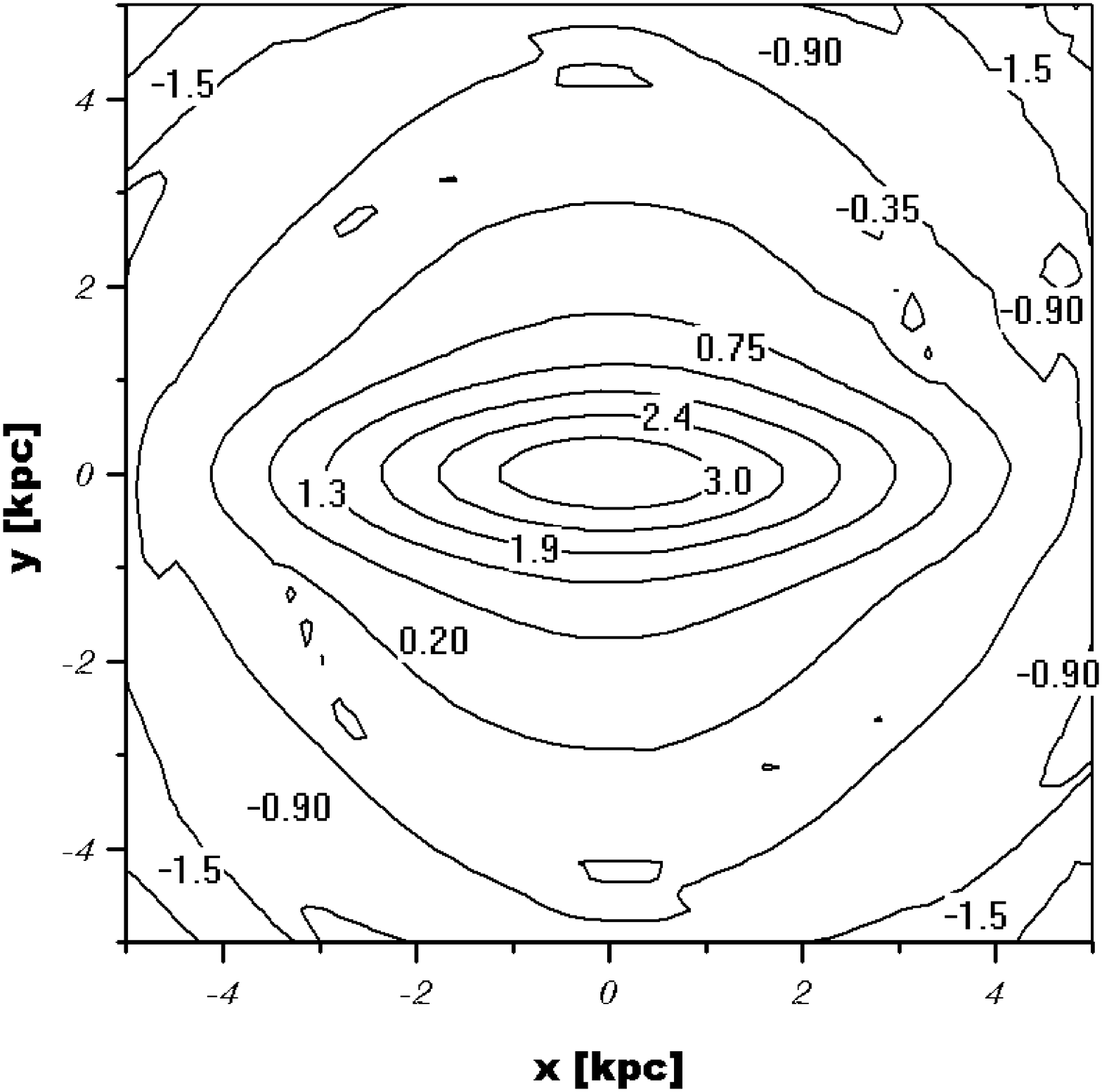}{}
  \caption{Projection of our model 20A onto the $xy$-plane. 
    To avoid modifying the bar contours by the strong, nearly axially
    symmetric disk component, only the density at $|z|> 225\pc$ was
    integrated. The length of the bar is $\simeq 3.5 {\rm \kpc}$ and its
    elongation is $\simeq $10:3-4.  Contours are in logarithmic surface
    density, with relative contour values indicated on the plot.}
\label{fignoebene}
\end{figure}

\begin{figure}
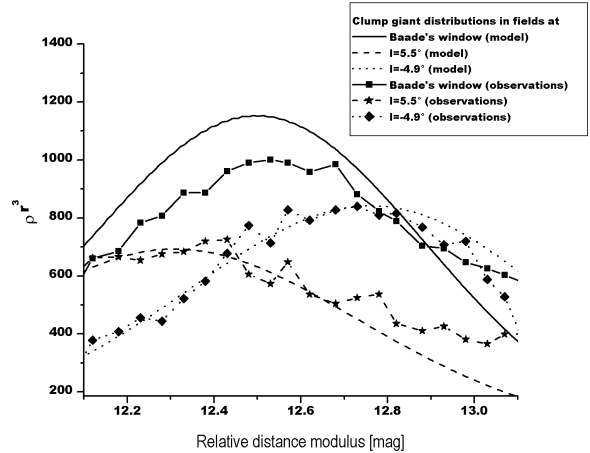

  \getfig{modellklumpen.eps2}{}
  \caption{Line-of-sight distributions of clump giants in the
    directions of three fields observed by Stanek \etal (1994,
    1997), 
    with symbols as given in the figure legend.  The abscissa is
    their $V_{V-I}$. For
    field M7 the observed counts were averaged over two CCD frames of
    equal angular size; for field M5 one CCD frame and for Baade's
    window six frames were used.  The curves show our best-fit model
    including spiral arms.  For the model, two constants were adjusted
    by eye as follows: (i) The model distributions were shifted along
    the abscissa such that the $l=5.5\deg$ and $l=-4.9\deg$ peaks
    match best the locations of the observed peaks.  (ii) The
    normalisation of the model curves was determined such as to
    approximately match the normalisation of the observed
    distributions in the fields M5 ($l\approx-4.9\deg$) and M7
    ($l\approx5.5\deg$).  The resulting shift and normalisation are
    then applied to all model distributions simultaneously.  All model
    distributions are convolved with an assumed width of the clump
    giant intrinsic luminosity distribution of $0.3^m$. 
}
  \label{figcgswith}
\end{figure}

\subsection{Models with bar but without spiral arms}
\label{sectnospiral}
To find a model for the {\sl COBE/DIRBE} L-band data without
spiral arms, we start the iterations from the
parametric model given by Binney, Gerhard \& Spergel \cite{bgs97},
and set the spiral structure penalty term in the
likelihood function to zero.  Fig.~\ref{figsbmaps} shows the surface
brightness map of this non-parametric model compared with the {\sl
  COBE/DIRBE} L-band data.  It fits the {\sl COBE} data well; the
iterations were stopped when the RMS of the model reached $0.073^m$.
On the left side of Fig.~\ref{figmodel} we show two cuts through this
model, in the upper panel a cut in the $xy$-(main) plane, and in
the lower panel a cut parallel to the $xz$-plane, at
$y\!\approx\! 85\pc$.  In the $xy$ map we can see overdensities
that point from the observer (at $x\!\approx\! 7.5\kpc$ and
$y\!\approx\! 2.7\kpc$) towards the tangential points of the spiral
arms, at approximately $l=\pm 30\deg$ and $l=-50\deg$ (see also
Drimmel \& Spergel \cite{drimmsp01}).

Can we find a model of the MW without spiral arms which fits the {\sl
  COBE/DIRBE} L-band data as well as this model, and does not have
such features?  To try to eliminate these overdensities, we have
computed models with larger penalty parameters for deviations from
eight-fold symmetry and/or smoothness.  However, we have only found
models which either contain similar features in the directions to the
spiral arm tangent points and achieve a ``good'' model RMS $\approx
0.07^m$, or models which are smooth and nearly eight-fold symmetric
without such features, but which then fit the data badly (model RMS
$\gta 0.2^m$).  Smooth models which fit the data well without these
features do not seem to exist. This suggests that a spiral arm
component is implied by the data. I.e., the luminosity in these
features is real, but the shift towards the observer is due to the FTS
effect discussed in \S\ref{sectperform}.

We mention that a similar effect was observed by Binney, Gerhard \&
Spergel \cite{bgs97} in their Richardson-Lucy (RL) models. They used
the same data of Spergel et al., and in their models found symmetric
density enhancements at $2-3 {\rm \kpc}$ down the minor axis of the
bar. They suggested that these features might be caused by spiral arms
being symmetrized by the RL algorithm, which forces the models to be
eight-fold symmetric.  In contrast to the RL models our deprojection
algorithm favours changes to the model density near the observer, and
so in the model discussed here the spiral arm overdensities are placed
near the observer. The projected bar elongation in our model without
spiral arms is $0.56$ ($|z|>225\pc$), comparable to that in the model of 
Binney, Gerhard \& Spergel \cite{bgs97}.

\subsection{Models with bar and spiral arms}
\label{sectbarmod}

We have seen that models for the MW L-band luminosity density develop
overdensities near the observer, towards the tangential directions of
Galactic spiral arms, when started from triaxially symmetric initial
distributions.  Thus we now proceed with a non-parametric density
estimation of the data by a model which includes spiral arms.  We show
that the inclusion of spiral structure not only improves the model for
the {\sl COBE/DIRBE} L-band data, but also results in a better match
to the {\los} distributions of clump giants towards certain bulge
fields. We also show that the derived structure of the bulge/bar does
not depend significantly on whether the assumed spiral model is two or
four-armed. 

First we fit a parametric model to the {\sl COBE/DIRBE} L-band data,
continuing to assume a bar angle $\phibar=20\deg$. This model has the
following bulge parameters (see \S\ref{secparametric}): $\eta=0.31$,
$\zeta=0.38$, $a_m=2.8\kpc$, $a_0=0.1\kpc$, $\rho^0_b=1180\, {\rm
  CLU}$, bar angle $\phibar=20\deg$; disk parameters:
$R_d=2.2\kpc$, $\alpha=0.65$, $z_0=0.19\kpc$, $z_1=0.042\kpc$,
$\rho^0_d=0.54\, {\rm CLU}/\kpc^3$; and spiral arm amplitude: $d_s=0.90$, for
the four-armed logarithmic spiral arm model similar to that of Ortiz \&
L\'epine \cite{ortiz93}. Here CLU are {\sl COBE} luminosity units as 
in Binney \etal \cite{bgs97}.

Starting the algorithm from this configuration and including the
spiral structure penalty term, we find a best non-parametric density
model. This model was selected from a
number of calculations run to fine-tune the penalty function
parameters, optimizing both the RMS and the model smoothness.  This
model, stopped at RMS $\approx 0.079^m$,
is one of our best if not the best model, and will be used as
reference model ``20A'' in what follows, deferring the discussion of
the acceptable range of $\phibar$ to Section \ref{barangle} below.
The RMS for this model is similar to that for the model without spiral
arms; the difference is not significant.  Correspondingly, both models
look very similar on the sky and match the data well; see Figure
\ref{figsbmaps}.  The main difference is in the three-dimensional
structure: Figure \ref{figmodel} shows cuts through both models. In
the model with spiral arms, these arms provide most of the
non-axisymmetric density.  Some residual luminosity is still required
towards some of the arm tangent points, but deviations from
point-symmetry in the Galactic plane near the Sun ($x\geq 3\kpc$,
$|y|\le 5\kpc$) are reduced by a factor of $\simeq 4.5$ (NB: the spiral
arms remain nearly point-symmetric during the iterations).

The quality of the model fit to the {\sl COBE/DIRBE} L-band data,
especially for the non-axisymmetric bar/bulge, is visualized by the
asymmetry maps shown in Figure \ref{figasymmap}. These maps show the
difference between the logarithmic fluxes at positive and negative
longitudes, ${\rm SB}(|l|,b)-{\rm SB}(-|l|,b)$ for both our reference model 20A
and the {\sl COBE/DIRBE} L-band data.  Generally, the model is
smoother than the data, but it reproduces the main bar-related
features of the observed map well. Note the good recovery of the
regions with clear bar signature around $(l,b)\approx (8\deg,\pm
5\deg)$, and the change of sign of the asymmetry near the galactic
centre. Here the far side of the bar appears brighter, a signature of
a bar with its near end at positive longitudes (Blitz \& Spergel
\cite{blitz91}). The most obvious differences between both maps are in
a strong feature at $(l,b)\approx (14\deg, 0\deg)$ in the observed
map, which may be local, and in the contours near zero asymmetry,
which are most affected by noise.

Figure \ref{fignoebene} shows the density of model 20A projected along
the $z$-axis for $|z|> 225\pc$. The density near the main plane of the
MW is excluded to avoid modifying the bar contours by the strong,
nearly axially symmetric inner disk component.  In model 20A the bar
is more elongated than in the model without spiral arms.  This is
because for the relative geometry of the arms, the bar, and the
position of the Sun, the spiral arms make the model appear broader in
longitude on the sky, and for fixed observed asymmetry this allows
the bar to be more elongated in the plane. The projection also
stresses the true extent of the bar, which is $\simeq 3.5\kpc$. The 
measured projected bar elongation in the $xy$-plane is $\simeq$10:3-4.
The contrast in the total face-on surface density 
between $(x,y)=(2.5\kpc,0)$ and
$(x,y)=(0, 2.5\kpc)$ is a factor of $\simeq 1.6$. A fit to the disk
profile in the radial range between $3.5\kpc$ and $5.5\kpc$ gives an
exponential radial scale-length of $2.1\kpc$.

We now compare the three-dimensional structure of the model to the
observations of Stanek \etal (\cite{stanek94}, \cite{stanek97}). 
These authors determined
the line-of-sight distributions of clump giant stars for a number of
fields towards the bulge/bar. Because these stars are of nearly
identical absolute magnitude within a small dispersion, measuring
their brightness distribution at a certain position on the sky
provides a profile of their density along the respective {\los}. 
The apparent magnitude of the peak of the distribution
locates the highest-density point along the {\los}.  The difference in
the apparent magnitude of the peak between various bulge fields,
especially between the fields at $l\approx5.5\deg$ and
$l\approx-4.9\deg$, reflects the shape of the non-axisymmetric
bar/bulge density distribution. These data therefore provide an
independent and strong test for the {\sl COBE/DIRBE} density models.
Note that the clump giant density maxima along the {\los} are insensitive
to a possible radial gradient in the ratio of clump giant stars to 
L-band luminosity.

For the comparison we fold the model {\los} density distributions with
a gaussian $\exp -({\Delta{\rm mag}^2}/{2 \sigma^2})$, to simulate the intrinsic
dispersion of clump giant absolute magnitudes.  In the literature
values $ 0.1^m\lta\sigma\lta0.3^m$ have been proposed (Stanek \etal
\cite{stanek97}, Stanek \& Garnavich \cite{stanekgarnavich},
Paczynski \& Stanek \cite{paczstan}). See Fig.~3 of Perryman \etal
\cite{perry1997} for an impression of the sharpness of the clump in
$V$ and Udalski \cite{udalski00} for an analysis of the metallicity
dependence on the mean I-band brightness.  We have explored a number
of different values for $\sigma$ between $0.2^m$ and $0.4^m$, and
finally decided for $\sigma=0.3^m$, because with this value our models
reproduce best the observations. For each model we need to select two
additional parameters, the normalisation of the model density and a
shift in magnitudes.  These represent the (unknown) conversion factor
between model density units and the number density of clump giant
stars, and the absolute magnitudes of clump giants.
For model 20A we determine these two parameters such that they fit
best the observations at $l\approx5.5\deg$ (field M7) and
$l\approx-4.9\deg$ (field M5).

Figure \ref{figcgswith} shows that the {\los} distributions of model
20A compare well with the clump giant observations of Stanek et al. 
Fitting Gaussians to the upper parts of the model curves yields an
asymmetry of $0.44^m$ between the {\los} at $l=-4.9\deg$ and
$l=5.5\deg$, even somewhat larger than observed.  Also the relative
peak heights and approximate widths of the model distributions agree
with the data within $\sim 10\%$. These are less of a constraint on the
bulge shape, however, because they are influenced by other parameters
like the density concentration of the bulge and the clump width
$\sigma$.  We remark that the choice of model normalisation factor
such that the main difference is in the peak height of Baade's window
distribution is arbitrary; we could also have decided to make a
near-perfect fit to Baade's window distribution and a $10\%$ error in
the peak heights of the other two distributions.  I.e., the model is
slightly more centrally concentrated than the clump giant
distribution.

\begin{figure*}
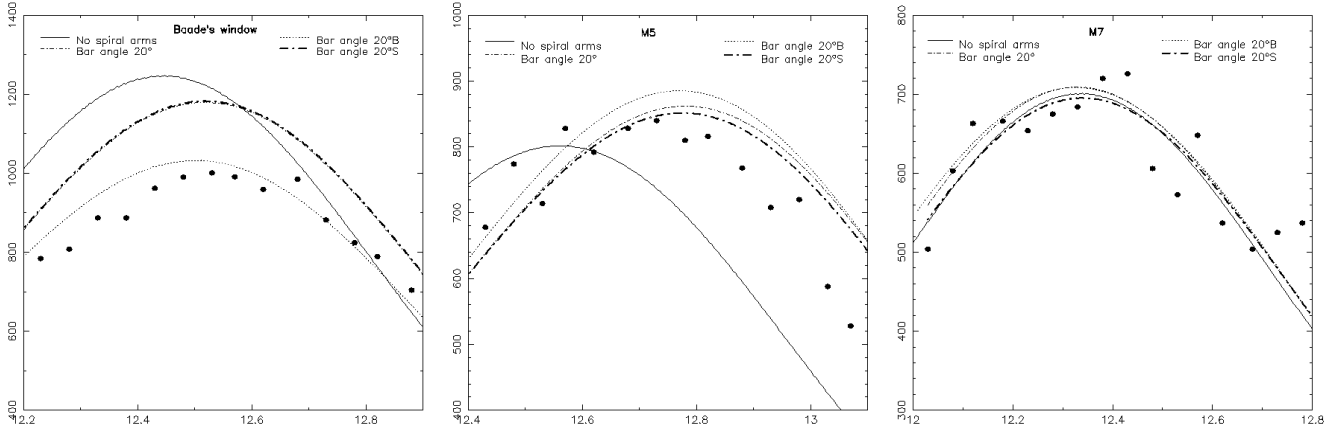

  \getfig{20degclumps.eps2}{}
  \caption{Comparison of observed clump giant line-of-sight distributions
    with several models, all for $\phibar=20\deg$: the model without spiral
    arms of \S\ref{sectnospiral} (full curve), our reference model 20A
    with spiral arms (dot-dashed), a similar model with broad spiral
    arms (20B, dotted), and a smoothed version of the reference model
    (20S, thick dot-dashed).}
\label{fig20degclumps}
\end{figure*}

The measured asymmetry in the new model is significantly larger than
in the eightfold-symmetric Richardson-Lucy models of Binney \etal
\cite{bgs97} and Bissantz \etal \cite{bissantz97}. These models have a
maximal asymmetry $\approx 0.27^m$, and generally $\leq 0.2^m$,
compared to the Stanek \etal \cite{stanek94} result of $0.37^m\pm
0.025^m$.  As Figure \ref{fig20degclumps} shows, it is also
significantly larger than the asymmetry in the model without spiral
arms from Section \ref{sectnospiral}, which is not a good fit to the
clump giant data.  In Fig.~\ref{fig20degclumps}, the magnitude scale
for the different models has been chosen so that they all match the
observed distribution in field M7. The smaller asymmetry of the model
without spiral arms thus becomes apparent as deviations in the peak
positions in BW and, in particular, field M5.  The spiral model 20A
has a greater asymmetry in the peak positions for these fields because
the elongation of its bar is larger, as discussed above.

Fig.~\ref{fig20degclumps} also shows a model ``20B'', which was
obtained by deprojecting the {\sl COBE} data with a modified broad spiral
arm model of FWHM $500\pc$, and a model ``20S'', which is a smoothed
version of the standard model with bar angle $20\deg$. For the
smoothing we have used the algorithm described in Section \ref{secseq}
in the form that preserves the spiral arms. Model 20S shows that
small-scale structure in the luminosity model does not influence the
model clump giant distributions in these bulge fields significantly.
Model 20B is actually a better fit to the amplitudes of the observed
distributions than our reference model 20A.  However, it does not fit
the L-band data as well (see Section \ref{barangle} below).

So far we have considered luminosity models in which the spiral arm
component, if present, has a four-armed structure. This is based
mainly on observations of gas tracers (see Vall\'ee \cite{vallee} and
Englmaier \& Gerhard \cite{ppe99}). However, it is unclear whether the
MW has two or four stellar spiral arms.  In the L-band data the
tangent point at $l\approx50\deg$ is not visible. This may point to
a two-armed structure; however, this tangent point is also weak in CO,
possibly due to the geometry of the {\los} through this arm (Dame, 
private communication).  Drimmel \& Spergel \cite{drimmsp01} argue
that the Sagittarius-Carina arm is - at least - weaker than the other
arms.  Therefore we now ask whether our results on the structure of
the bar/bulge depend on the assumption of a four-armed spiral model.
We have generated two non-parametric models of the {\sl COBE/DIRBE} L-band
data in which a two-armed parametric model was used both for the
starting model and the spiral arm penalty term. The bar angle is still
assumed to be $20\deg$. In the first model, 
the arms start near the major axis of the bar,
and the pitch angle is half that used in the four-armed model above.
In the second, we omit the Sagittarius-Carina arm and its counter-arm,
the arms start near the minor axis of the bar, and the pitch angle is the
same as in the four-armed model. 
Compared with the four-armed model 20A, both
two-armed models found by the algorithm show only minor differences.
They fit the {\sl COBE/DIRBE} L-band data equally well as model 20A; the
asymmetry in the clump giant {\los} distribution peak positions
differs by $\lta 0.03^m$, the peak heights differ by $\lta 12\%$, and
the elongation of the bar/bulge differs by $\lta 4\%$. 
It appears therefore that the Scutum-Crux arm is most important for the
deprojection of the bar. Thus the
assumption of a four-armed spiral model does not significantly bias
the results obtained for the structure of the bar/bulge.

We end this section by a short summary of its main results.  The first
is that inclusion of spiral structure improves the model of the {\sl
  COBE/DIRBE} L-band data in the sense of removing unphysical
structures in the disk plane.  Second, in models including spiral
structure the bar is more elongated as compared to triaxially
symmetric models, and third, this more elongated bar gives a better
representation of the observed apparent magnitude distributions of
clump giant stars in several bulge fields.

\begin{figure}
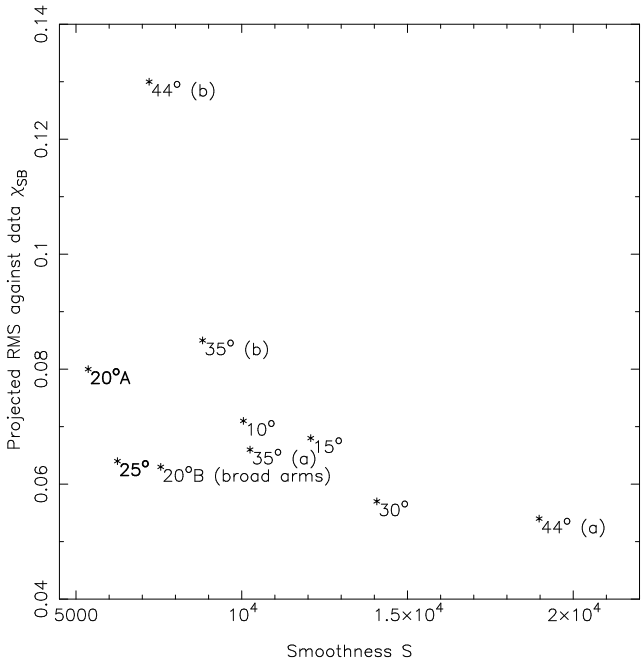

  \getfig{phi_chiintrplot.eps2}{}
  \caption{Smoothness parameter $S$ and model RMS 
    $\chi_{\rm SB}$ for non-parametric models with spiral arms
    of the {\sl COBE/DIRBE}
    L-band data, for bar angles $10\deg\leq \phibar\leq44\deg$. Models
    in the lower left of the diagram both provide a good fit to the
    surface brightness and are the smoothest.  Those printed in the
    diagram in bold face are acceptable models in the sense of
    $F\leq3$; cf.\ Section \ref{secseq}.  Model 20B for bar angle
    $20\deg$ and with broad spiral arms is the only ``marginally
    acceptable'' model ($3<F\leq4$).  Thus bar angles
    $20\deg\leq\phibar\leq25\deg$ are preferred.}
  \label{fig_phichiintr}
\end{figure}    

\begin{figure*}\begin{center}
\psfig{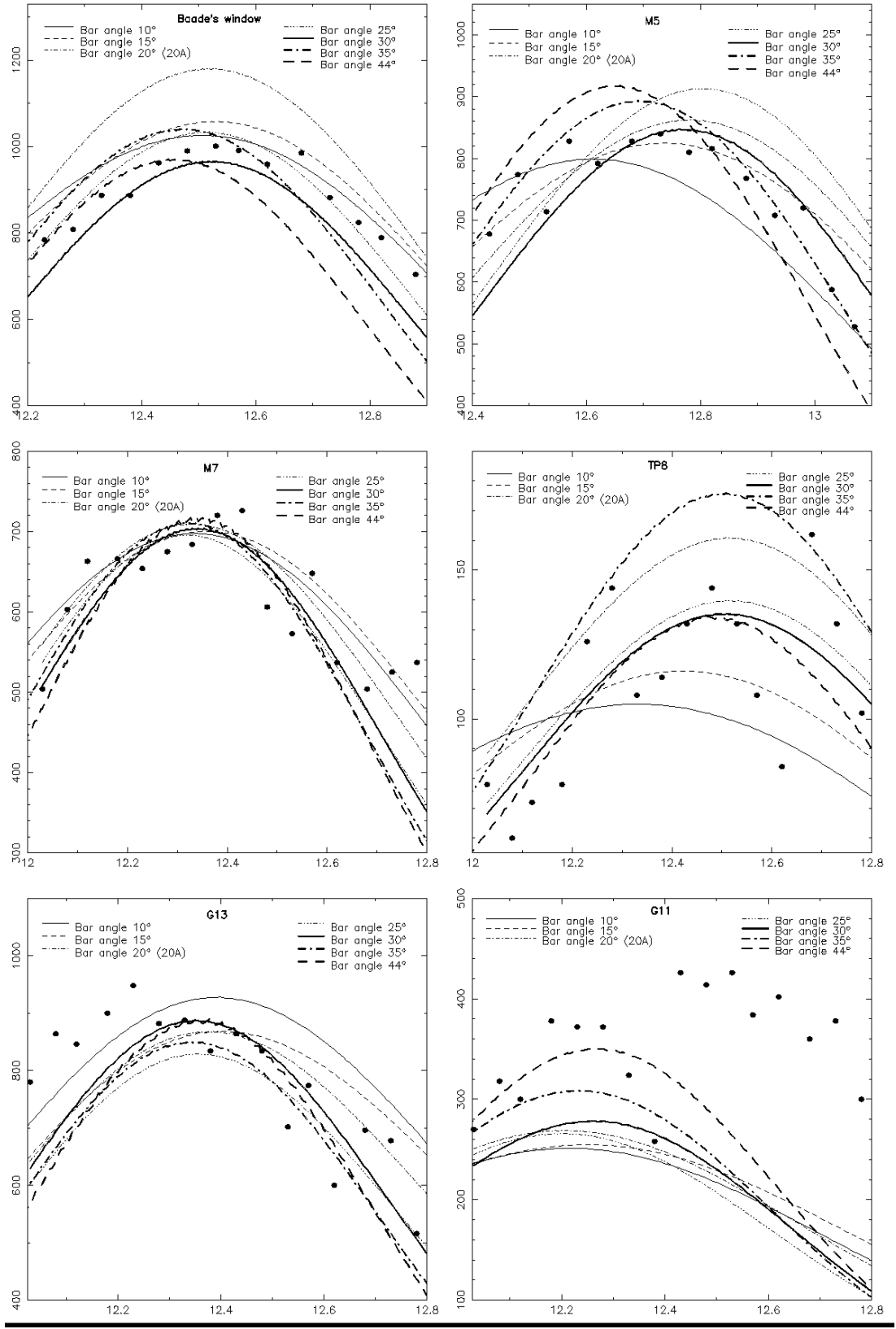}\end{center}
  \caption{Clump giant distributions for models with different bar angles.
For each model, the normalization and shift in the magnitude
scale are determined such that the model fits best the data in field M7.
The best fit is found for bar angles $\phibar\approx 15\deg-30\deg$.}
\label{figresclump}
\end{figure*}    

\subsection{Constraining the bar angle}
\label{barangle}
In the previous section we have described model 20A for bar angle
$\phibar=20\deg$ as our reference model. We will now construct similar
models for a variety of bar angles, and use them to constrain the
possible range of $\phibar$. This will also make clear why we selected
model 20A.  We will compare these models in three different ways.  In
the first (and weakest) test we use the quality of fit of the {\sl
  COBE} data for the best-fit {\sl parametric} models used as starting
models in the non-parametric deprojections. In the second test we
compare the non-parametric models themselves, using the $F$-criterium
of Section \ref{secseq} which measures a combination of quality of
fit to the data and model smoothness. Finally, the third test uses the
distribution of clump giant apparent magnitudes along the {\los}
measured by Stanek \etal (\cite{stanek94}, \cite{stanek97}).

We have non-parametrically estimated the {\sl COBE} data for bar
angles $\phibar=10\deg,15\deg,20\deg,25\deg,30\deg,35\deg,44\deg$,
using the standard four-armed spiral model, and additionally for
$\phibar=20\deg$ using a model with broad spiral arms of FWHM
$500\pc$. In each case we started the iterations from a corresponding
parametric best-fit model.

First we analyse these parametric initial models, and find that they
fit best the part of the sky dominated by the central bulge/bar,
around $|l|\!\approx\! 8\deg$, $b\!\approx\! \pm 5\deg$, when the
bar angle is $20\deg\leq\phibar\leq 30\deg$.  For other bar angles the
models show systematic deviations from the {\sl COBE/DIRBE} L-band
data in these regions in $(l,b)$, symmetric to the galactic equator.

Second we compare the non-parametric 
models using the $F$ criterion (goodness of
fit combined with smoothness), as introduced in Section \ref{secseq}.
Their smoothness parameters $S$ and model RMS are shown in Figure
\ref{fig_phichiintr}.  Models in the lower left corner give the best
fit to the surface brightness and have the highest degree of
smoothness.  Acceptable models ($F\leq3$) are the standard model 20A
($\phibar=20\deg$), and the $\phibar=25\deg$ model. The
$\phibar=20\deg$ model with broad spiral arms is marginally acceptable
($3<F\leq4$). The other models are not satisfactory: they are either
too unsmooth or do not fit the SB data well. We illustrate the
trade-off between goodness of fit and smoothness in these cases with
two models for $\phibar=35\deg$ and $\phibar=44\deg$, obtained with
different smoothness penalty parameters $\lambda$.  One of these is
clearly not smooth, and the other is smooth but fits the {\sl COBE}
data poorly.  We conclude that bar angles $20\deg\leq\phibar\leq
25\deg$ are preferred.

Finally, we compare the predicted clump giant line-of-sight
distributions of these models with observations by Stanek \etal
(\cite{stanek94}, \cite{stanek97}) in Figure \ref{figresclump}.  The
intrinsic dispersion of clump giant absolute magnitudes is again set
to $\sigma=0.3^m$ (see Section \ref{sectbarmod}).  The remaining free
parameters of the models, that is the normalization and the magnitude
shift, are fixed by optimizing the model fit to the observations in
field M7.  These parameters are then identical for all fields.
 
Fig. \ref{figresclump} shows that several models fit the observed
clump giant line-of-sight distributions nearly equally well. None of
the models fits the observations in field G11
(at $(l,b)\approx(8.2\deg,-4.4\deg)$), probably since these
data are strongly influenced by the underlying broad population of
stars (the power-law part in the luminosity function fitted by Stanek
\etal).  The models with $\phibar=10\deg$ and $\phibar=44\deg$ provide
inferior fits to the data. The $\phibar=10\deg$ model has wrong peak
positions, heights, and widths for fields TP8 
(at $(l,b)\approx(-0.1\deg,-8.0\deg)$)
and M5, and in Baade's
window the peak width is too large. The $\phibar=44\deg$ model shows
wrong peak positions in Baade's window and field M5, and a deficit in
asymmetry between Baade's window/field M5 and field M7 (for this model
the fit can be improved slightly by using a very high intrisic
dispersion $\sigma\!\gta\! 0.4^m$. However, this is far above
published values, see \S\ref{sectbarmod}). The other models with
$15\deg\lta\phibar\lta30\deg$ cannot be distinguished on the basis of
the present data. 

\begin{figure*}
  \getfig{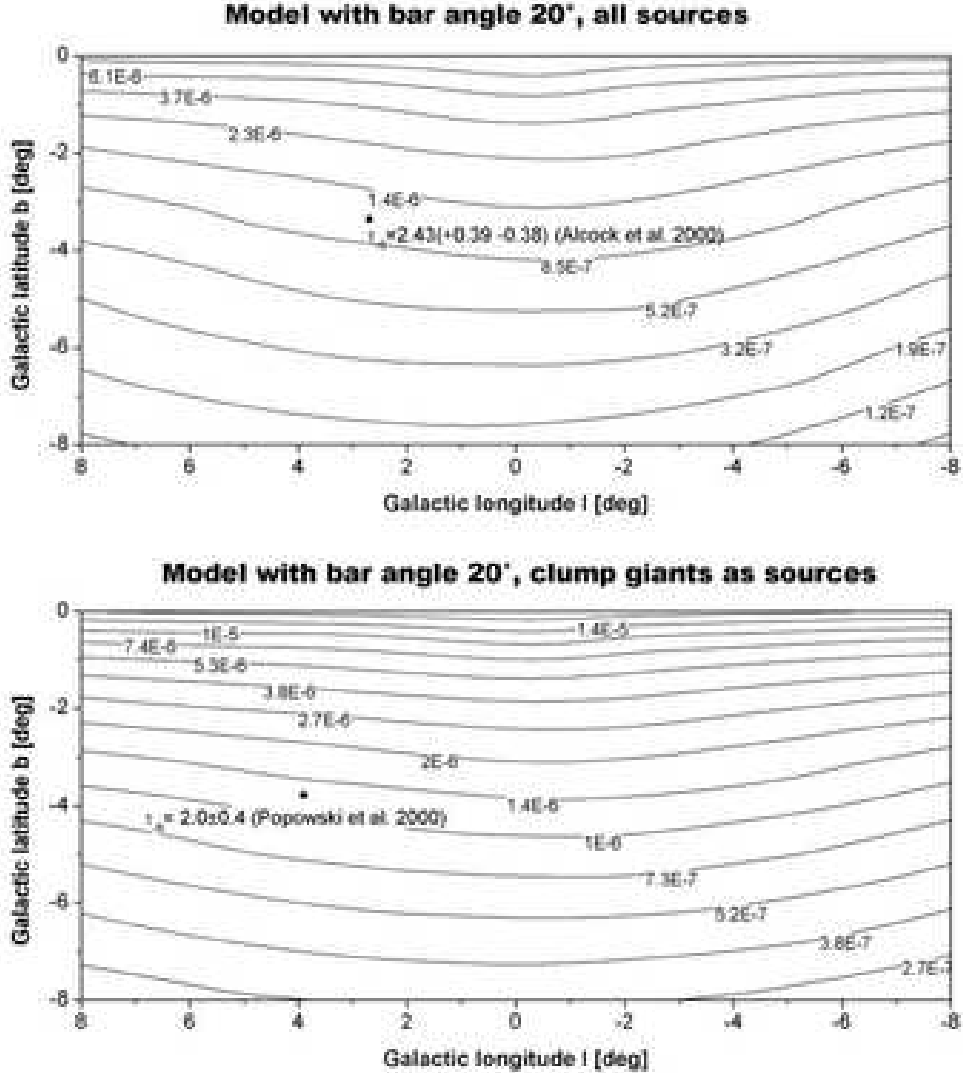}{}
  \caption{Microlensing optical depth map of our reference model 
   including spiral arms. The model is for bar angle $20\deg$. The upper 
   map shows the optical depth
   for all sources, the lower map for clump giant sources only. The
   mean positions of the newly published {\sl MACHO} results are indicated in 
   each map.}
  \label{figmlmap}
\end{figure*}

In summary, a bar angle $20\deg\leq\phibar\leq25\deg$ is consistent
with all three tests. The clump giant data are consistent with a wider
range, $15\deg\leq\phibar\leq30\deg$; however, for $\phibar=15\deg$
and $\phibar=30\deg$ we have not been able to find models passing the
F-criterion (goodness of fit combined with smoothness).  The
$20\deg$ model 20A stands out by its smoothness.

\section{Microlensing}

We will now provide predictions for the microlensing optical depth for
our NIR models. The required conversion factor from luminosity density
to mass density can be found from fitting the observed terminal
velocity curve with a model for the gas flow in the gravitational
potential of the bar and disk, assuming spatially constant L-band
mass-to-light ratio $M/L_L$ (Bissantz et al. \cite{bissantz97},
Englmaier \& Gerhard \cite{ppe99}).  In a forthcoming paper
(Bissantz, Englmaier \& Gerhard, in preparation) we will describe SPH
simulations of the gasdynamics in the potentials of the new luminosity
models presented in the present work.  The value of $M/L_L$ derived
from the terminal curve depends somewhat on the precise model
parameters, for example, the pattern speed. For the best SPH model it
is $M/L_L=3.9 \times 10^8 M_\odot / $ CLU, and between the various
models it varies in the range $3.7$-$4.1 \times 10^8 M_\odot / $CLU,
i.e., by $\pm 5\%$. In the following we will use $M/L_L=3.9 \times
10^8 M_\odot / $ CLU.

With this value, the optical depth towards Baade's window (BW)
$(l\!=\! 1\deg, b\!=\! -3.9\deg)$ for our reference model 20A is
$\tau_{-6}=0.95$ for the full sample of source stars, using $\beta=-1$
in the parametrization of Kiraga \& Paczynski \cite{kp94} in
accounting for a magnitude cut-off.  For clump giant sources only it
is $\tau_{-6}=1.39$ in BW ($\beta=0$).  In Figure \ref{figmlmap} we
present optical depth maps for both cases, predicted from model 20A,
over the entire inner Galaxy region.  At constant mass normalisation,
the range in luminosity density through BW predicted by models 15,
20B, 30 corresponds to an uncertainty in $\tau$ of about 10 percent.

\begin{figure}
  \getfig{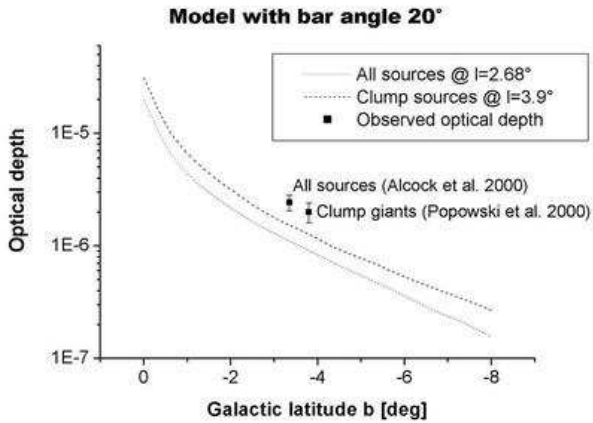}{}
  \caption{Microlensing optical depth of our reference model at the 
  longitudes of the newly
  published {\sl MACHO} results, plotted as function of galactic latitude. The 
  observations are indicated in the figure. The upper curve shows the optical
  depth for clump giant sources, the lower curve for all sources. Both curves are
  for the galactic longitude of the published observations for the respective
  group of sources.}  
  \label{figmlprofil}
\end{figure}

The {\sl MACHO} group has published revised values for the optical
depth near Baade's window based on their new bulge microlensing data.
They give an optical depth for all sources, based on $99$ events from
3 yr of data, of $\tau_{-6} = 2.43^{+0.39}_{-0.38}$ averaged over 8
fields centred at $l=2.68\deg$ and $b=-3.35\deg$ (Alcock et al.
\cite{alcock2000a}). From $52$ microlensing events with clump giant
sources in 5 yr of data, Popowski et al. \cite{popo2000} give
$\tau_{-6}=2.0\pm 0.4$ at a mean position $l=3.9\deg$ and
$b=-3.8\deg$. The corresponding values predicted from model 20A are
$\tau_{-6}=1.10$ at $l=2.68\deg$ and $b=-3.35\deg$ ($\beta=-1$), and
$\tau_{-6}=1.27$ for clump giant sources at $l=3.9\deg$ and
$b=-3.8\deg$ ($\beta=0$).  Figure \ref{figmlprofil} shows profiles of
optical depth along galactic latitude at the mean longitudes of the
MACHO observations. The curve for $l\!=\!2.68\deg$ shows the optical
depth for all sources, and that for $l\!=\!3.9\deg$ the optical depth
for clump giant sources only, for comparison with the observational
results. Both curves illustrate the steep dependence of $\tau$ with
Galactic latitude.

The optical depth in the new NIR models is about $10\%$ higher than
for the eight-fold symmetric RL models of Bissantz et al.
\cite{bissantz97}. This near agreement between two independent
non-parametric models for the {\sl COBE/DIRBE} data is gratifying.
What difference there is mostly comes from the $10\%$ higher
luminosity to mass conversion ($M/L_L$) for the new model. The effects
of the structural differences in the new model appear to
compensate. On the one hand, there are more lens stars in the new
model where the line-of-sight to Baade's window crosses the spiral
arms, increasing $\tau$.  On the other hand, because the total surface
density along this {\los } is approximately constant (since specified
by the {\sl DIRBE} SB), the density in the inner bulge is lower than
in the models of Bissantz et al. \cite{bissantz97}. This decreases
the average distance to the sources, and hence $\tau$.

Compared to the observed optical depths, the predictions of the new
model are still low.  For clump giant sources only, the model is
consistent with the preliminary new value (Popowski \etal
\cite{popo2000}) $\tau_{-6}=2.0\pm 0.4$ to within $1.8\sigma$.  One
assumption we have made is that the microlensed source stars are
distributed similar to the luminous matter in the Galaxy.  This is
confirmed by the good agreement of the clump giant distributions
predicted from the NIR model with those measured by Stanek \etal
(\cite{stanek94}, \cite{stanek97}). Further evidence that the lensed
stars do not contain a significant component far behind the Galactic
center (e.g., in the Sagittarius dwarf) comes from the CMD in Fig.~2
of Popowski \etal. With the distribution of source stars known, the
predicted optical depth can only be modified significantly if the
distribution of lenses is substantially different from that of the
sources, i.e., if mass does not follow NIR light.

Associated uncertainties in the NIR model prediction were discussed by
Bissantz \etal \cite{bissantz97}. There appears to be two main causes for
concern: (i) The corrections by Spergel \etal \cite{spergel96} for
dust absorption might conceivably have caused us to overestimate the
luminosity in the Galactic plane. In this case, there could be room
for some lensing dark mass in front of the bulge fields. (ii)
Independently, the L-band mass-to-light ratio might vary with position
in the inner Galaxy.  Both would have the effect of modifying the mass
distribution of the inner Galaxy but, as discussed by Bissantz et al., the
effect of this on the optical depth is limited to $\sim 20\%$ because
of the constraints from the terminal velocity curve. Together with the
$10\%$ spread in the model optical depth discussed above,
this implies an uncertainty in the predicted clump giant value of
order 0.3, i.e., there is no strong discrepancy with the clump giant
measurement of Popowski \etal.

However, the high optical depth of $\tau_{-6} = 2.43^{+0.39}_{-0.38}$
for all sources measured from difference imaging analysis (DIA, Alcock
\etal \cite{alcock2000a}) is $3.5\sigma$ away from the predicted value
of model 20A ($\tau_{-6} = 1.10$), and even after allowing for a
$30\%$ uncertainty in the predicted optical depth is still more than
$2.5\sigma$ discrepant. From the measured optical depth, Alcock \etal
deduced $3.23^{+0.52}_{-0.50}$ for bulge sources only, assuming a 25\%
contribution from disk sources.  Binney, Bissantz, \& Gerhard \cite{bbg00}
have shown such high optical depths cannot plausibly be reconciled
with the Galactic rotation curve and the mass density near the Sun.
To underline their argument, to increase the optical depth from
$\tau_{-6} = 1.10$ to $\tau_{-6} = 2.43$ for the same distribution of
sources and $\beta=-1$, would require an additional surface mass
density even at near-optimal distance, $\simeq4\kpc$, of some $1540
\msun/\pc^2$, comparable to the luminous mass density already present
in the NIR mass model ($3636\msun/\pc^2$).  This may suggest a problem
in the interpretation of the DIA measurement, for example, in the
correction for amplification bias.

We end this section by commenting on the microlensing contribution of
the MW's dark halo. The NIR models with the quoted $M/L_L$ reproduce
the Galactic terminal velocity 
curve out to $\gta 5\kpc$ without inclusion of a
dark halo (Bissantz, Englmaier \& Gerhard, in preparation).  If the
LSR circular speed is $v_c=220{\rm km/s}$ (consistent with
$R_0=8\kpc$, Reid \etal \cite{reid99}, Backer \& Sramek
\cite{backer99}), some dark matter is required between $5\kpc$ and the
solar radius, but most of this will be at high latitudes, while the
line-of-sight to Baade's window, for example, is within one disk
scaleheight $z_0$ in this range of Galactocentric radii. Decreasing
the amount of luminous mass in the inner Galaxy in favour of dark
matter also does not help even if the dark matter microlenses; this
case is included in the $\sim 20\%$ uncertainty discussed
above. Moreover, from the LMC microlensing results (Alcock \etal
\cite{alcock2000b}) we know that at most a small fraction of this dark
matter would actually microlense, so that this would likely {\sl
decrease} the predicted optical depth towards the bulge.

\section{Summary and conclusions}

We have developed a penalized maximum likelihood algorithm that
enables us to non-parametrically estimate luminosity densities from
surface brightness data. In testing this algorithm with artificial
data we found a degeneracy in the bar shape for fixed bar angle, which
essentially comes from noise in the data.  This implies the existence
of a sequence of parametric and non-parametric models that within a
given noise level in the data can not be distinguished. For the noise
level typical of the {\sl COBE/DIRBE} data, this sequence corresponds
to an uncertainty of the in-plane bar elongation $\eta$ of $\pm0.1$
and a corresponding variation in the half mass radius $r_{0.5}$ of the
bar/bulge of $\pm20\%$.  However, we show that the degeneracy between
models on this sequence can be broken by comparing them with the
line-of-sight distributions of clump giant stars.

We have non-parametrically estimated luminosity distributions for the
{\sl COBE} data, including a model for spiral structure in the disk.
This is done in two steps. First we fit a parametric model to the
data, then we improve this with the non-parametric algorithm.  The
initial model contains a spiral arm term proposed by Ortiz \& L\'epine
\cite{ortiz93}, which is also used as a prior in a penalty term that is added
to the likelihood function in subsequent iterations.  Models with
spiral arms do not have the unrealistic finger-to-Sun features that
are commonly seen in models with an axisymmetric disk, and at the same
time fit the surface brightness maps equally well.

We have considered a sequence of models with varying bar angles
$10\deg\leq\phibar\leq44 \deg$. We evaluate these using both a
criterion measuring a combination of the goodness-of-fit to the {\sl
  COBE} data and the intrinsic smoothness of the luminosity
distribution, and the degree to which they account for the asymmetry
in the clump giant line-of-sight distributions from Stanek \etal
(\cite{stanek94}, \cite{stanek97}). In this way we find a preferred
range $15\deg\lta\phibar\lta30\deg$, with the best models found for
$20\deg\lta\phibar\lta25\deg$.  In our reference $\phibar=20\deg$
model, the length of the bar is $\approx 3.5\kpc$, and its axis ratios
are $10:3-4:3$. The in-plane elongation is larger than in previous
eightfold symmetric luminosity distributions, because spiral arms make
the model appear broader on the sky, thereby requiring a more
elongated bar for fixed surface brightness data.  The more elongated
bar in turn increases the asymmetry in the peak distances of the
model's clump giant line-of-sight distributions in the fields observed
by Stanek \etal (\cite{stanek94}, \cite{stanek97}), enabling the new 
model to reproduce these observations well.

Analysing a model with two spiral arms instead of the four-armed
structure of Ortiz \& L\'epine \cite{ortiz93}, we have concluded that our
results regarding the structure of the bar/bulge structure and the fit
to the clump giant line-of-sight observations do not depend
significantly on the assumed spiral arm model, as long as the spiral
arm tangent points as seen from the Sun are similar.

The microlensing optical depth in Baade's window for our reference model is
$\tau_{-6} \approx0.95$ for all sources and $\tau_{-6}\approx 1.39$ for
clump giant sources only, when the NIR mass-to-light ratio is assumed to
be constant and is determined by fitting to the Galactic terminal
velocity curve (maximal disk model, 
Bissantz, Englmaier \& Gerhard, in preparation).
For clump giant sources at $(l,b)=(3.9\deg,-3.8\deg)$ we find
$\tau_{-6}\equiv\tau/10^{-6}=1.27$, within $1.8\sigma$ of the new
MACHO measurement $\tau_{-6}=2.0\pm 0.4$ 
given by Popowski et al. \cite{popo2000}. The value for all sources
at $(l,b)=(2.68\deg,-3.35\deg)$ is $\tau_{-6}=1.1$, still $>3\sigma$
away from the published MACHO DIA value $\tau_{-6}=2.43^{+0.39}_{-0.38}$. 
The dispersion of these
$\tau_{-6}$ values within our models is $\simeq 10\%$. Because the NIR
model is a good representation for the distribution of microlensing 
sources, the predicted values can only be modified significantly if the
distribution of lenses is different from that of the sources. This, however,
is constrained because of the good fit of the predicted model terminal
curve to the Galactic terminal curve. As we have previously estimated
(Bissantz \etal \cite{bissantz97}), this makes it difficult to increase
the predicted optical depths by $>20\%$. 
Thus the MW disk and bulge must have near-maximal mass-to-light ratio
to explain even the clump giant value for the optical depth.
As Binney \etal \cite{bbg00}
have argued, optical depths as high as the DIA value are difficult
to obtain by any model that is constrained by the Galactic rotation
curve and local disk density.

\end{document}